\documentclass[11pt]{article}
\usepackage{amsmath,amssymb}
\usepackage{graphicx,color}
\usepackage{braket}

\numberwithin{equation}{section}

\def\bra#1{\langle#1|}
\def\ket#1{|#1\rangle}

\def\({\left(}
\def\){\right)}

\newcommand{\de}{\partial}
\newcommand{\be}{\begin{equation}}
\newcommand{\ba}{\begin{eqnarray}}
\newcommand{\ea}{\end{eqnarray}}
\newcommand{\ee}{\end{equation}}

\newcommand{\f}{\frac}
\newcommand{\s}{\sqrt}
\newcommand{\vp}{\varphi}

\newcommand{\ti}{\tilde}
\newcommand{\ap}{\alpha}

\newcommand{\ddd}{\cdot\cdot\cdot}
\newcommand{\no}{\nonumber \\}
\newcommand{\la}{\langle}
\newcommand{\lb}{\rangle}
\newcommand{\ep}{\epsilon}

 \def\al{\alpha'}
 \def\de{\partial}

 \def\f {\frac}
 \def\ti{\tilde}
 \def\ap{\alpha}

 \def\ddd{\cdot\cdot\cdot}
 \def\no{\nonumber \\}
 \def\la{\langle}
 \def\lb{\rangle}
 \def\ep{\epsilon}

\newcommand{\bal}{\begin{align}}
\newcommand{\eal}{\end{align}}
\newcommand{\bes}{\begin{equation*}}
\newcommand{\bas}{\begin{eqnarray*}}
\newcommand{\eas}{\end{eqnarray*}}
\newcommand{\ees}{\end{equation*}}
\newcommand{\bals}{\begin{align*}}
\newcommand{\eals}{\end{align*}}

\newcommand{\p}{\partial}

\begin{document}

\begin{titlepage}
\thispagestyle{empty}

\begin{flushright}
YITP-14-18\\
IPMU-14-0049\\
IPM/P-2014/010\\
\end{flushright}

\vspace{.4cm}
\begin{center}
\noindent{\Large \textbf{Entanglement between Two Interacting CFTs \\
and \\
Generalized
Holographic Entanglement Entropy}}\\
\vspace{2cm}

Ali Mollabashi $^{a,b}$,
Noburo Shiba $^{b}$,
 and
Tadashi Takayanagi $^{b,c}$
\vspace{1cm}

{\it
$^{a}$School of physics, Institute for Research in Fundamental Sciences (IPM), Tehran, Iran\\
 $^{b}$Yukawa Institute for Theoretical Physics (YITP),\\
Kyoto University, Kyoto 606-8502, Japan\\
$^{c}$Kavli Institute for the Physics and Mathematics of the Universe (Kavli IPMU),\\
University of Tokyo, Kashiwa, Chiba 277-8582, Japan\\
}

\vskip 2em
\end{center}

\vspace{.5cm}
\begin{abstract}
In this paper we discuss behaviors of entanglement entropy between two interacting CFTs
and its holographic interpretation using the AdS/CFT correspondence. We explicitly perform analytical calculations of entanglement entropy between two free scalar field theories which are interacting with each other in both static and time-dependent ways. We also conjecture a holographic calculation of entanglement entropy between two interacting ${\cal N}=4$ super Yang-Mills theories by introducing a minimal surface in the S$^5$ direction, instead of the AdS$_5$ direction. This offers a possible generalization of holographic entanglement entropy.
\end{abstract}

\end{titlepage}

\newpage


\newpage

\section{Introduction}

\hspace{5mm} Quantum entanglement offers us a useful tool to study global properties of quantum field theories (QFTs). In particular, one of the most important quantities for this purpose is the entanglement entropy $S_A$ which measures the amount of
quantum entanglement between a subsystem $A$ and its complement $B$.

Indeed this quantity captures basic structures of any given QFT.
For example, this quantity follows the area law \cite{Bombelli:1986rw,Sr,Ereview,Lreview} if we consider a local quantum field theory with a UV fixed point, while non-local field theories \cite{ShTa} or QFTs with fermi surfaces \cite{FS} at UV cut off scale can violate the area law. Moreover, the coefficients of logarithmically divergent terms of $S_A$ in even dimensional conformal field theories (CFTs) are proportional to central charges \cite{HLW,CC,RT,MS,Sol}. Thus $S_A$ can detect degrees of freedom of CFTs. It is also useful to note that the entanglement entropy can quantify topological properties in gapped systems \cite{TEE}.

In most of literature, the entanglement entropy is geometrically defined by separating the spatial manifold into the subsystem $A$ and $B$. Instead, the main purpose of this paper is to analyze entanglement entropy between two CFTs (called CFT$_1$ and CFT$_2$) which live in a common spacetime and interacting with each other, described by the action of the form:
\be
S=\int dx^{d}\left[{\mathcal L}_{CFT_1}+{\mathcal L}_{CFT_2}+{\mathcal L}_{int}\right].
\ee
Since the total Hilbert space is decomposed as the direct product:
\be
{\mathcal H}_{tot}={\mathcal H}_{CFT_1}\otimes {\mathcal H}_{CFT_2},
\ee
we can define the entanglement entropy between CFT$_1$ and CFT$_2$ by tracing out the total density matrix $\rho_{tot}$ over either of them:
\be
S_{ent}=-\mbox{Tr}\rho_1\log \rho_1, \ \ \ \rho_1=\mbox{Tr}_{{\mathcal H}_{CFT_2}}[\rho_{tot}]. \label{twoee}
\ee
Note that we can exchange CFT$_1$ with CFT$_2$ in the above definition as long as
the total system is pure. It is also obvious that if there are no interactions between them, $S_{ent}$ gets vanishing. Thus this entanglement entropy may offer us a universal measure of strength of interactions.

 Such a problem was already analyzed in \cite{FK,Xu,CF,LFFO} mainly from condensed matter viewpoints. In \cite{FK,LFFO} and \cite{CF}, the entanglement entropy between two coupled Tomonaga-Luttinger liquids and Heisenberg antiferromagnets was computed, respectively.
 In \cite{Xu}, the behavior of entanglement entropy between two CFTs in the presence of interacting perturbations was studied. In our paper, we will present analytical results by focusing on a solvable relativistic example: two copies of a massless free scalar field theory which are defined in any dimension and are interacting with each other at any order of relativistic interactions. We will present two different but equivalent methods of calculations: (i) a real time formalism based on wave functionals and (ii) an Euclidean replica formalism using boundary states. Owing to these methods, we will furthermore study time evolutions of entanglement entropy when we turn on interactions instantaneously at a time.

In the light of AdS/CFT correspondence \cite{Maldacena}, the geometries of gravitational spacetimes can be encoded in the quantum entanglement of dual CFTs as is expected from the holographic calculation of entanglement entropy \cite{RT,HRT}. Therefore the AdS/CFT correspondence relates the global structures of gravitational spacetimes to those of CFTs in an interesting way
(see e.g. \cite{MERA,Mark,BLR,HuRa,CKNR,BHMERA,cMERA,BiMy}). This consideration raises one interesting question. In string theory examples of AdS/CFT, a gravity dual usually includes an internal compact space in addition to the AdS spacetime, as is typical in the AdS$_5\times$ S$^5$ type IIB string background dual to the four dimensional ${\mathcal N=4}$ super Yang-Mills. Therefore one may wonder how quantum entanglement in CFTs
can probe internal spaces such as S$^5$. As we will discuss in the final part of this paper, this question is closely related to the entanglement entropy between two interacting CFTs (\ref{twoee}). This should be distinguished from a system of two entangling CFTs without interactions, where its gravity dual is given by an AdS black hole geometry \cite{EBH}.

This paper is organized as follows: In section two we will give an explicit and analytical calculation of entanglement entropy between two massless scalar fields in the real time formalism based on wave functionals. In section three, we will provide an alternative calculation based on Euclidean replica formalism using boundary states and analyze time evolutions of entanglement entropy. In section four, we will discuss its holographic counterpart and conjecture a generalization of holographic entanglement entropy. In section five, we summarize our results and discuss future problems.

\section{Entanglement between Two Interacting CFTs}

 In this section we introduce our relativistic scalar field models of two interacting CFTs and perform analytical computations of (both von-Neumann and Renyi) entanglement entropy in these setups based on the direct real time formalism in terms of the wave functionals. Refer to the paper \cite{Xu} for an earlier interesting analysis of related problems for Renyi entropy using a perturbation theory and scaling argument. Refer also to \cite{FK,LFFO} for a field theoretic analysis of Tomonaga-Luttinger liquids.
 Here we will give explicit non-perturbative results in a solvable relativistic QFT with several choices of interactions, which also allows us to calculate of its time evolutions.

\subsection{Our Models}
We consider two models of two interacting QFTs.
The first model is the massless interaction model whose action is given by,
\begin{equation}
S=\dfrac{1}{2} \int d^d x [ (\partial_\mu \phi)^2
+ (\partial_\mu \psi)^2 +\lambda \partial_\mu \phi   \partial^\mu \psi ]  .  \label{conformal action}
\end{equation}
Interestingly, the equation of motion is $\partial^2 \phi =\partial^2 \psi=0$ which	 does not depend on $\lambda$.
However, the Hamiltonian and the conjugate momenta depends on $\lambda$ and
the nonzero $\lambda$ causes entanglement in the ground state.
We diagonalize the action by the orthogonal transformation,
\begin{equation}
S=\dfrac{1}{2} \int d^d x [A_{+} (\partial_\mu \phi')^2
+A_{-} (\partial_\mu \psi')^2 ]  ,  \label{daction}
\end{equation}
where 
\begin{equation}
A_{\pm}=1\pm \tfrac{\lambda}{2}  ,  \label{Apm}
\end{equation}
and
\begin{equation}
\begin{pmatrix}
\phi '  \\
\psi  '
\end{pmatrix}    =\dfrac{1}{\sqrt{2}}
\begin{pmatrix}
1 && 1  \\
-1 && 1
\end{pmatrix}
\begin{pmatrix}
\phi   \\
\psi
\end{pmatrix}  .     \label{conformal transformation}
\end{equation}
From (\ref{daction}) we obtain the Hamiltonian,
\begin{equation}
H=\dfrac{1}{2} \int d^{d-1} x \left[A_{+}^{-1} \left( \pi_{\phi'}^2 +A_{+}^2 (\nabla \phi')^2\right)
+ A_{-}^{-1} \left( \pi_{\psi'}^2 +A_{-}^2 (\nabla \psi')^2\right) \right]  ,  \label{conformal hamiltonian}
\end{equation}
where $ \pi_{\phi'}$ and  $ \pi_{\psi'}$ are the conjugate momenta of $\phi'$ and $\psi'$.
From (\ref{Apm}) and (\ref{conformal hamiltonian}) we see that the positivity of the Hamiltonian
 restricts the range of $\lambda$ as
\begin{equation}
-2 \leq \lambda \leq 2  .  \label{lambda}
\end{equation}

The second model is the massive interaction model whose action is given by,
\begin{equation}
S=\dfrac{1}{2} \int d^d x \left[ (\partial_\mu \phi)^2
+ (\partial_\mu \psi)^2
-
(\phi , \psi )
\begin{pmatrix}
A & C  \\
C & B
\end{pmatrix}
\begin{pmatrix}
\phi   \\
\psi
\end{pmatrix} \right]
 ,  \label{massive action}
\end{equation}
where $A, B$ and  $C$  are real constants whose dimensions are $(mass)^2$.
We diagonalize the action by the orthogonal transformation,
\begin{equation}
S=\dfrac{1}{2} \int d^d x \left[ (\partial_\mu \phi ')^2
+ (\partial_\mu \psi ')^2
- m_1^2 \phi'^2 - m_2^2 \psi'^2
 \right]
 ,  \label{massive daction}
\end{equation}
where
\begin{equation}
\begin{pmatrix}
\phi '  \\
\psi  '
\end{pmatrix}    =
\begin{pmatrix}
\cos \theta & -\sin \theta  \\
\sin \theta & \cos \theta
\end{pmatrix}
\begin{pmatrix}
\phi   \\
\psi
\end{pmatrix}  .     \label{massive transformation}
\end{equation}
Note that the three parameters $(A,B,C)$ in (\ref{massive action}) are equivalently expressed in terms of $(\theta,m_1,m_2)$ in (\ref{massive daction}). For the stability of our model, we requires $m_1\geq 0$ and $m_2\geq 0$. The parameter $\theta$ expresses the strength of interaction except when $m_1=m_2$.

We will trace out $\phi$ and consider the entanglement entropy of $\psi$ in the following
two cases. First, we consider the entanglement entropy of the ground state for the total Hamiltonian.
Next, we consider the time evolution of the entanglement entropy
generated by the total Hamiltonian.
We choose the initial state to be the ground state for the free Hamiltonian which is the Hamiltonian of
free massless fields,
i.e. we prepare the ground state for the free Hamiltonian and switch on the interaction at $t=0$.


\subsection{Entanglement entropy for the gaussian wave function}
Because the Hamiltonians of our models are quadratic, the wave functions of the ground states and of
time-evolving states whose initial states are the ground states for the free Hamiltonian are
gaussian wave functions.
In this section we consider generally the entanglement entropy for the gaussian wave function.
We can calculate the entanglement entropy by the similar method to \cite{Callan:1994py},
where geometric entropy for the ground state of a free scalar field which is
a real valued gaussian wave function was considered.

We consider the following gaussian wave function,
\begin{equation}
\begin{split}
\braket{ \{ \phi , \psi \}  | \Psi} = &
N \exp \bigg\{-\dfrac{1}{2} \int d^{d-1} x d^{d-1} y \Big[ \phi(x) G_1(x,y) \phi (y) +
 \psi (x)G_2(x,y) \psi (y)   \\
& +2 \phi(x) G_3(x,y) \psi (y)\Big] \bigg\}
 .  \label{general wavefunction}
\end{split}
\end{equation}
where $N$ is a normalization constant and $G_i(x,y)$ ($i=1,2,3$) are the complex valued functions and $G_i(x,y)=G_i(y,x)$.

We trace out $\phi$ and obtain the density matrix of $\psi$ as
\begin{equation}
\begin{split}
&\rho_{\psi}(\psi_1,\psi_2) = \int D \phi  \braket{ \{ \phi , \psi_1 \}  | \Psi }  \braket{\Psi | \{ \phi , \psi_2 \} } \\
& =N' \exp \left[-\dfrac{1}{2}  \int d^{d-1} x d^{d-1} y
 (\psi_1(x),\psi_2(x))
\begin{pmatrix}
             X(x,y) & 2Y(x,y)  \\
             2Y(x,y) & X^\ast(x,y)
             \end{pmatrix}
\begin{pmatrix}
             \psi_1(y)  \\
             \psi_2(y)
             \end{pmatrix} \right] ,
\end{split}
  \label{general density}
\end{equation}
where $N'$ is a normalization constant and
\begin{equation}
X=G_2-G_3 (G_1+G_1^\ast)^{-1} G_3 , ~~~
Y=\dfrac{-1}{4} \left[G_3 (G_1+G_1^\ast)^{-1} G_3^\ast +G_3^\ast (G_1+G_1^\ast)^{-1} G_3 \right].
\label{general XY}
\end{equation}
In (\ref{general XY}) we have considered $G_i$ as symmetric matrices with continuous indices $x,y$
and the products are the products of matrices.
From the normalization condition we obtain
\begin{equation}
1=\mbox{Tr} \rho_\psi =N' [\mathrm{det} (\pi^{-1} (\mathrm{Re}X+2Y))]^{-1/2},
 \label{general normalization}
\end{equation}
From (\ref{general density}) we obtain
\begin{equation}
\mbox{Tr}\rho_\psi^n =
N'^n \int D\psi_1 \cdots D\psi_n \exp \left[ -\int d^{d-1} x d^{d-1} y  (\psi_1(x), \cdots, \psi_n(x)) M_n (x,y)
\begin{pmatrix}
             \psi_1(y)  \\
            \vdots \\
             \psi_n(y)
             \end{pmatrix} \right] \label{general trace}
\end{equation}
where
\begin{equation}
M_n =
\begin{pmatrix}
            \mathrm{Re}X & Y & 0 & \cdots &0 &Y   \\
            Y& \mathrm{Re}X&Y&\cdots &0&0 \\
             0&Y& \mathrm{Re}X&\cdots &0&0 \\
            \vdots &\vdots &\vdots & &  \vdots &\vdots \\
            0&0&0&\cdots &\mathrm{Re}X&Y \\
             Y&0&0&\cdots &Y&\mathrm{Re}X \\
             \end{pmatrix} . \label{general M_n}
\end{equation}
From (\ref{general normalization}) and (\ref{general trace}) we obtain
\begin{equation}
\mbox{Tr} \rho_\psi^n =\left[\mathrm{det}\left(\pi^{-1}(\mathrm{Re}X+2Y)\right)\right]^{n/2} \left[\mathrm{det}\left(\pi^{-1}M_n \right)\right]^{-1/2} .
\label{general trace2}
\end{equation}
We rewrite $M_n$ as
\begin{equation}
 M_n = \dfrac{\mathrm{Re}X}{2} \tilde{M_n}, \label{M_n2}
\end{equation}
where
\begin{equation}
\tilde{M_n} =
\begin{pmatrix}
            2 & -Z & 0 & \cdots &0 &-Z   \\
            -Z&2&-Z&\cdots &0&0 \\
             0&-Z&2&\cdots &0&0 \\
            \vdots &\vdots &\vdots & &  \vdots &\vdots \\
            0&0&0&\cdots &2&-Z \\
             -Z&0&0&\cdots &-Z&2 \\
             \end{pmatrix} , \label{eq:9}
\end{equation}
here
\begin{equation}
Z=-2\left(\mathrm{Re}X\right)^{-1} Y =-2 \left[Y^{-1} \mathrm{Re} G_2 +2+ Y^{-1}
\mathrm{Im} G_3 (\mathrm{Re}G_1)^{-1} \mathrm{Im}G_3\right]^{-1}  . \label{general z}
\end{equation}
We diagonalize $Z$ and denote the eigenvalues of $Z$ as $z_i$.
And we can diagonalize $\tilde{M_n}$ by Fourier transformation and obtain
\begin{equation}
\mathrm{det} \tilde{M_n} =
\prod_i \prod_{r=1}^n \left[2-2z_i \cos \left(\dfrac{2 \pi r}{n}\right) \right] =\prod_i 2^n \dfrac{(1-\xi_i^n)^2}{(1+\xi_i^2)^n},  \label{general detM_ntilde}
\end{equation}
where $\xi_i$ is defined as
\begin{equation}
z_i=\f{2\xi_i}{(\xi_i^2+1)}. \label{xi}
\end{equation}
From (\ref{xi}) we obtain  the solution $\xi$ as
\begin{equation}
\xi_i=\dfrac{1}{z_i} \left[1-\sqrt{1-z_i^2}\right] . \label{massive xi 2}
\end{equation}

From (\ref{general trace2}) and (\ref{general detM_ntilde}) we obtain
\begin{equation}
\mbox{Tr}\rho_\psi^n = \prod_{i} \dfrac{(1-\xi_i)^n }{(1-\xi_i^n)} .
\label{massive trace3}
\end{equation}
Finally we obtain the Renyi entropies $S_\psi^{(n)}=(1-n)^{-1} \ln \mbox{Tr} \rho_\psi^n$
and the entanglement entropy
$S_{\psi}=-\mbox{Tr} \rho_\psi \ln \rho_\psi =-\tfrac{\partial}{\partial n}
\ln \mbox{Tr} \rho_\psi^n |_{n=1} $
as follows:
\begin{align}
&S_\psi^{(n)}   \equiv \sum_i s^{(n)}(\xi_i)=
\sum_i (1-n)^{-1} \left[n \ln (1-\xi_i)-\ln(1-\xi_i^n) \right] ,     \label{general r entropy} \\
&S_\psi   \equiv \sum_i s(\xi_i)=
\sum_i \left[- \ln (1-\xi_i)-\dfrac{\xi_i}{1-\xi_i} \ln \xi_i \right]
.\label{general entropy}
\end{align}

Because the entanglement entropy is a complicated function of $G_i$,
we will expand it.
For later convenience, we expand $s^{(n)}(\xi)$ and $s(\xi)$ as functions of $z$.
For $z\ll1$, we obtain
\begin{equation}
s^{(n)}(\xi)\approx (1-n)^{-1}\left[-n\dfrac{z}{2} +\left(\dfrac{z}{2}\right)^n\right], ~~~~ s(\xi)\approx -\dfrac{z}{2} \ln z.
\label{entropy small z}
\end{equation}
For $1- z\ll1$, we obtain
\begin{equation}
s^{(n)}(\xi) \approx -\dfrac{1}{2}\ln (1-z), ~~~~ s(\xi)\approx -\dfrac{1}{2}\ln (1-z).
\label{entropy large z}
\end{equation}


\subsection{Massless Interactions}
We apply the above formalism to the massless interaction case (\ref{conformal action}).
\subsubsection{Ground states}
First we compute the ground state wave function.
From (\ref{conformal hamiltonian}) we obtain the ground state wave function as
(see e.g. the equation (7) in \cite{Bombelli:1986rw})
\begin{equation}
\braket{ \{ \phi' , \psi' \}  | \Omega} =
N \exp \left\{-\dfrac{1}{2} \int d^{d-1} x d^{d-1} y W(x,y)\left[A_{+} \phi'(x) \phi' (y) +A_{-} \psi'(x) \psi' (y) \right] \right\}
 ,  \label{dwavefunction}
\end{equation}
where $N$ is a normalization constant and
\begin{equation}
W(x,y)=V^{-1} \sum_{k} w e^{ik(x-y)}.
\end{equation}
Here $w=|k|$ and $V$ is the volume of the space and we impose the periodic boundary condition.
Note that $\ket{\phi',\psi'} =\ket{\phi,\psi}$ because  $\{ \phi ,\psi \}$ is the orthonormal transformation of $\{ \phi', \psi' \}$.
Thus we rewrite the ground state wave function by $\{ \phi ,\psi \}$ and obtain
\begin{equation}
\braket{ \{ \phi , \psi \}  | \Omega} =
N \exp \left\{-\dfrac{1}{2} \int d^{d-1} x d^{d-1} y W(x,y) \left[ \phi(x) \phi (y) + \psi (x) \psi (y) +\lambda  \phi(x) \psi (y)  \right] \right\}
 .  \label{wavefunction}
\end{equation}
Because this is a gaussian wave function,
we can obtain the Renyi entropies $S_{1}^{(n)}$
and the entanglement entropy $S_{1}$ from 
(\ref{general r entropy}) and (\ref{general entropy}) as
\begin{align}
&S_{1}^{(n)} = s_{1}^{(n)} (\lambda)\cdot
\sum_{k\neq0}1,     \label{massless r entropy} \\
&S_{1}  =s_{1}(\lambda)\cdot
\sum_{k\neq0}1
 ,\label{massless entropy}
\end{align}
where
\begin{align}
&s_{1}^{(n)} =
(1-n)^{-1} [n \ln (1-\xi)-\ln(1-\xi^n) ] ,     \label{massless r entropy 2} \\
&s_{1}  =
- \ln (1-\xi)-\dfrac{\xi}{1-\xi} \ln \xi
 .\label{massless entropy 2}
\end{align}
here
\begin{equation}
\xi=\dfrac{1}{z} \left[1-\sqrt{1-z^2}\right] ,~~~ z=\dfrac{\lambda^2}{8-\lambda^2}.
\label{massless xi}
\end{equation}

\begin{figure}[ttt]
   \begin{center}
     \includegraphics[height=4cm]{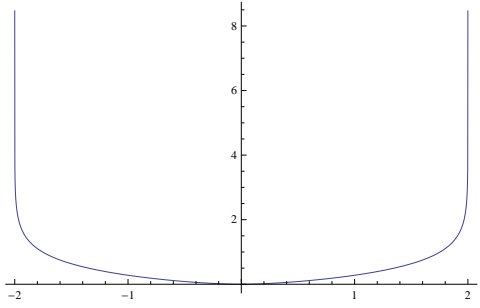}
   \end{center}
   \caption{The plot of $s_1(\lambda)$, which is proportional to the entanglement entropy due to the massless interaction for the range $-2<\lambda<2$..}\label{fig:sla}
\end{figure}

In this case each mode contributes identically the entropies.
The contribution to the entropy from the zero mode is zero.
The mode sum is UV divergent and we regularize it by the smooth momentum cutoff.
The profile of $s_1(\lambda)$ is plotted in Fig.\ref{fig:sla}. It gets divergent at
$\lambda=\pm 2$ where the interaction becomes maximal.
In this way, it is obvious that the entanglement entropy between two scalar field theories
has the volume law divergence and its coefficient is a monotonically increasing function of the coupling constant $|\lambda|$.

When the volume of the space is finite, there is a UV finite term in entropy densities.
For the periodic boundary condition in $d=2$ (the space is $S^{1}$ whose circumference is $L$) we obtain
\begin{equation}
\sum_{k\neq 0}\exp [-\epsilon |k|]=2\sum_{n=1}^{\infty} \exp [ -\epsilon 2\pi n/L]
=\dfrac{L}{\pi \epsilon } -1 +O(\epsilon/L) , \label{casimir}
\end{equation}
where $\epsilon$ is the UV cutoff length.
From  (\ref{massless r entropy}) and (\ref{casimir}) we obtain for $d=2$
\begin{align}
&S_1^{(n)}|_{d=2} =s_{1}^{(n)} (\lambda) \left(\dfrac{L}{\pi \epsilon } -1 +O\left(\epsilon/L\right)\right)
\label{entprl} \\
&S_1|_{d=2}  =s_{1} (\lambda) \left(\dfrac{L}{\pi \epsilon } -1 +O\left(\epsilon/L\right)\right).\label{entprkl}
\end{align}

We can pick up the UV finite term by differentiating the entropy density (we take $\epsilon$ to zero)
\begin{align}
&\dfrac{\partial}{\partial L}(S_1^{(n)} /L)|_{d=2} =s_{1}^{(n)}(\lambda)/L^2, \\
&\dfrac{\partial}{\partial L}(S_1/L)|_{d=2}  =s_{1}(\lambda) /L^2.
\end{align}
This UV finite entropy is analogous to the Casimir energy and is universal.
When the space is a $(d-1)$ torus of size $L$, we obtain
\begin{align}
&S_{1}^{(n)} = s_{1}^{(n)} (\lambda)
 \left(c_{d,d-1} \left( \dfrac{L}{\epsilon} \right)^{d-1} +
c_{d,d-2} \left( \dfrac{L}{\epsilon} \right)^{d-2} +\cdots
+c_{d,0} \right),     \label{massless r entropy 3} \\
&S_{1}  =s_{1}(\lambda) \left(c_{d,d-1} \left( \dfrac{L}{\epsilon} \right)^{d-1} +
c_{d,d-2} \left( \dfrac{L}{\epsilon} \right)^{d-2} +\cdots
+c_{d,0} \right)
 ,\label{massless entropy 3}
\end{align}
where $c_{d,l}$ are constants and $c_{d,0}$ is universal.
The universal term depends on the shape of the space and the boundary condition
as the Casimir energy.

Finally we expand $s_1^{(n)}$ and $s_1$ as functions of $\lambda$.
From (\ref{entropy small z}) and (\ref{massless xi}), we obtain for $|\lambda| \ll1$,
\begin{align}
&s_1^{(n)}(\lambda) \simeq (1-n)^{-1} \left[ -\dfrac{n\lambda^2}{16} +\left( \dfrac{\lambda^2}{16} \right)^{n} \right] ,  \label{entml} \\
&s_1 (\lambda) \simeq -\dfrac{\lambda^2}{16} \ln (\lambda^2 ) .
\end{align}
From (\ref{entropy large z}) and (\ref{massless xi}), we obtain for $2-|\lambda| \ll1$,
\begin{align}
&s_1^{(n)}(\lambda) \simeq -\dfrac{1}{2} \ln (2-|\lambda|) , \\
&s_1 (\lambda) \simeq -\dfrac{1}{2} \ln (2-|\lambda|)  .
\end{align}

\subsubsection{Time evolution}
In order to obtain the wave function at $t$,
let us recall the propagator of one harmonic oscillator.
The Hamiltonian is given by,
\begin{equation}
H = \dfrac{a^2}{2} p^2 + \dfrac{b^2}{2} q^2,
\end{equation}
where $a$ and $b$ are real positive constants and $[q,p]=i$.
We obtain the propagator as
\begin{equation}
\bra{q_1} e^{-iHt} \ket{q_2}
= \sqrt{\dfrac{b}{i 2\pi a \sin (abt) }}
\exp \left[i \dfrac{b}{2a \sin (abt)} \{ (q_1^2 +q_2^2) \cos (abt) -2 q_1 q_2 \} \right] .
\end{equation}

Next we consider the following Hamiltonian in quantum field theory,
\begin{equation}
H = \int d^{d-1} x \left[ \dfrac{A^2}{2} \pi^2 + \dfrac{B^2}{2} ((\nabla \phi )^2 +m^2 \phi^2)\right] ,
\end{equation}
where  $A$ and $B$ are real positive constants and
$[\phi(x), \pi (y)]=i \delta^{d-1} (x-y)$.
As a generalization of the propagator in one harmonic oscillator,
we obtain the propagator as
\begin{equation}
\begin{split}
\bra{\phi_1} e^{-iHt} \ket{\phi_2}
=& N(t)\exp \bigg\{\dfrac{i}{2} \int d^{d-1} x \int d^{d-1}y\Big[\left(\phi _1 (x) \phi_1(y) + \phi_2 (x) \phi_2 (y) \right) W_{g1}(x,y) \\
& -2\phi_1 (x) \phi _2 (y) W_{g2}(x,y) \Big]\bigg\} , \label{general propagator}
\end{split}
\end{equation}
where $N(t)$ is a normalization constant and
\begin{equation}
\begin{split}
&W_{g1}(x,y)=V^{-1} \sum_{k} \dfrac{Bw \cos (ABwt)}{A\sin (ABwt)} e^{ik(x-y)} , \\
&W_{g2}(x,y)=V^{-1} \sum_{k} \dfrac{Bw }{A\sin (ABwt)} e^{ik(x-y)} ,
\end{split}
\end{equation}
here $w=\sqrt{k^2+m^2}$ and $V$ is the volume of the space and we impose the periodic boundary condition.

We use the above propagator to compute the wave function at $t$
in the massless coupling case (\ref{conformal action}).
From the Hamiltonian (\ref{conformal hamiltonian}) and (\ref{general propagator}),
we obtain
\begin{equation}
\begin{split}
&\bra{\phi'_1,\psi'_1} e^{-iHt} \ket{\phi'_2,\psi'_2}
= N(t)
\exp \bigg\{\dfrac{i}{2} \int d^{d-1} x \int d^{d-1}y
\Big[ \big(A_+ \left(\phi' _1 (x) \phi'_1(y) + \phi'_2 (x) \phi'_2 (y) \right) \\
&+A_- \left(\psi' _1 (x) \psi'_1(y) + \psi'_2 (x) \psi'_2 (y) \right)\big) W_1(x,y)
 -2\left(A_+\phi'_1 (x) \phi' _2 (y) +A_-\psi'_1 (x) \psi' _2 (y)\right) W_2(x,y) \Big]\bigg\} .
\end{split}
\end{equation}
where
\begin{equation}
\begin{split}
&W_1(x,y)=V^{-1} \sum_{k} \dfrac{w \cos (wt)}{\sin (wt)} e^{ik(x-y)} , \\
&W_2(x,y)=V^{-1} \sum_{k} \dfrac{w }{\sin (wt)} e^{ik(x-y)} ,
\end{split}
\end{equation}
here $w=|k|$.
Note that $\ket{\phi',\psi'} =\ket{\phi,\psi}$ because  $\{ \phi ,\psi \}$ is the orthonormal transformation of $\{ \phi', \psi' \}$.
We rewrite the propagator in terms of $\{ \phi ,\psi \}$ and obtain
\begin{equation}
\begin{split}
&\bra{\phi_1,\psi_1} e^{-iHt} \ket{\phi_2,\psi_2}
= N(t)
\exp \bigg\{\dfrac{i}{2} \int d^{d-1} x \int d^{d-1}y
\Big[ \big(\phi _1 (x) \phi_1(y) + \phi_2 (x) \phi_2 (y)  \\
&+ \psi _1 (x) \psi_1(y) + \psi_2 (x) \psi_2 (y)
+\lambda (\phi_1(x) \psi_1 (y)+\phi_2(x) \psi_2 (y))\big) W_1(x,y)  \\
& -2\left(\phi _1 (x) \phi_2(y) + \psi_1(x) \psi_2 (y)
+\dfrac{\lambda}{2}(\phi_1 (x) \psi _2 (y) +\psi_1 (x) \phi _2 (y)) \right) W_2(x,y) \Big] \bigg\} . \label{conformal propagator}
\end{split}
\end{equation}
When the initial state is $\ket{0}$ which is the ground state for $\lambda=0$,
the wave function is given by,
\begin{equation}
\bra{\phi_1,\psi_1} e^{-iHt} \ket{0} =
\int D \phi_2 D \psi_2  \bra{\phi_1,\psi_1} e^{-iHt} \ket{\phi_2,\psi_2}
\braket{\phi_2,\psi_2|0}.
\end{equation}
where
\begin{equation}
\braket{ \{ \phi_2 , \psi_2 \}  | 0} =
N \exp \left\{-\dfrac{1}{2} \int d^{d-1} x d^{d-1} y W(x,y) \left[ \phi_2(x) \phi_2 (y) + \psi_2(x) \psi_2 (y) \right] \right\}
,
\end{equation}
here
\begin{equation}
W(x,y)=V^{-1} \sum_{k} w e^{ik(x-y)}.
\end{equation}
We perform the gauss integral and obtain the wave function at $t$ as
\begin{equation}
\begin{split}
\bra{\phi_1,\psi_1} e^{-iHt} \ket{0}
&= N \exp \bigg\{-\dfrac{1}{2} \int d^{d-1} x d^{d-1} y \Big[ \phi_1(x) G_1(x,y) \phi_1 (y) +
 \psi_1 (x)G_1(x,y) \psi_1 (y)   \\
& +2 \phi_1(x) G_3(x,y) \psi_1 (y)  \Big] \bigg\}
 .
\end{split}
\end{equation}
where
\begin{equation}
G_1(x,y)=V^{-1} \sum_{k} w g_1(wt) e^{ik(x-y)}, ~~~ G_3(x,y)=V^{-1} \sum_{k}w g_3(wt) e^{ik(x-y)}
\end{equation}
and
\begin{align}
&g_1(x)=\dfrac{1}{e^{2ix} -\dfrac{\lambda^2}{4}\cos^2 x}
\left[i\left(1-\dfrac{\lambda^2}{4} \right) \cot x - \left(1+\dfrac{\lambda^2}{4} \right)\right] -i\cot x  \\
&g_3(x)=\dfrac{1}{e^{2ix} -\dfrac{\lambda^2}{4}\cos^2 x}
\left[i\left(1-\dfrac{\lambda^2}{4} \right) \dfrac{\lambda}{2} \cot x - \lambda \right]
-i \dfrac{\lambda}{2}\cot x .
\end{align}
We can obtain the Renyi entropies $S_{2}^{(n)}(t)$
and the entanglement entropy $S_{2}(t)$ for $\bra{\phi_1,\psi_1} e^{-iHt} \ket{0} $
 from 
(\ref{general r entropy}) and (\ref{general entropy}) as
\begin{align}
&S_2^{(n)} (t)  \equiv \sum_{k\neq 0} s_2^{(n)} (wt)=
\sum_{k \neq 0} (1-n)^{-1} \left[n \ln \left(1-\xi (wt)\right)-\ln\left(1-\xi^n (wt)\right) \right] ,    \\
&S_2 (t)   \equiv \sum_{k\neq0} s_2(wt)=
\sum_{k\neq 0} \left[- \ln \left(1-\xi(wt)\right)-\dfrac{\xi (wt)}{1-\xi(wt)} \ln \xi (wt) \right]
 .
\end{align}
where
\begin{equation}
\xi(x) =\dfrac{1}{z(x)} \left[1-\sqrt{1-z(x)^2}\right] .
\end{equation}
and
\begin{equation}
z= \left[2 |g_3|^{-2} \left(\mathrm{Re} g_1\right)^2 -1+2  |g_3|^{-2}\left(\mathrm{Im} g_3\right)^2 \right]^{-1}  .
\end{equation}
For $|\lambda|\ll 1$, we obtain
\begin{equation}
z(x)\simeq \dfrac{\lambda^2}{2} \sin^2x.
\end{equation}
Thus we find 
\begin{align}
&S_\psi^{(n)} \simeq -\dfrac{n}{1-n} \dfrac{\lambda^2}{4} \sum_{k\neq0} \sin^2 (wt)   ~~~~(n>1),    \label{timentropy} \\
&S_\psi \simeq -\dfrac{\lambda^2}{4} \ln \lambda^2  \sum_{k\neq0} \sin^2 (wt).
\end{align}
More detailed time-dependent behavior including a physical interpretation will be studied in the next section.

\subsection{Massive Interactions}
\subsubsection{Ground states}
From (\ref{massive daction}) we obtain the ground state wave function as
(see e.g. the equation (7) in \cite{Bombelli:1986rw})
\begin{equation}
\braket{ \{ \phi' , \psi' \}  | \Omega} =
N \exp \bigg\{-\dfrac{1}{2} \int d^{d-1} x d^{d-1} y  \left[ \phi'(x) W_1(x,y) \phi' (y) +
 \psi'(x) W_2(x,y) \psi' (y) \right] \bigg\}
 ,  \label{massive dwavefunction}
\end{equation}
where $N$ is a normalization constant,
\begin{equation}
W_{1,2}(x,y)=V^{-1} \sum_{k} (k^2+m_{1,2}^2)^{1/2} e^{ik(x-y)},
\end{equation}
and 
$V$ is the volume of the space and we impose the periodic boundary condition.
Note that again we have $\ket{\phi',\psi'} =\ket{\phi,\psi}$, since  $\{ \phi ,\psi \}$ is the orthonormal transformation of $\{ \phi', \psi' \}$.
Thus we rewrite the ground state wave function in terms $\{ \phi ,\psi \}$ and obtain
\begin{equation}
\begin{split}
\braket{ \{ \phi , \psi \}  | \Omega} = &
N \exp \bigg\{-\dfrac{1}{2} \int d^{d-1} x d^{d-1} y \Big[ \phi(x) G_1(x,y) \phi (y) +
 \psi (x)G_2(x,y) \psi (y)   \\
& +2 \phi(x) G_3(x,y) \psi (y)  \Big] \bigg\}
 .  \label{massive wavefunction}
\end{split}
\end{equation}
where
\begin{align}
&G_1(x,y)\equiv V^{-1} \sum_k G_1(k) e^{ik(x-y)}
= V^{-1} \sum_k \left[\cos^2\theta \left(k^2+m_1^2\right)^{1/2} +
\sin^2 \theta \left(k^2+m_2^2\right)^{1/2} \right]e^{ik(x-y)} \\
&G_2(x,y)\equiv V^{-1} \sum_k G_2(k) e^{ik(x-y)}
= V^{-1} \sum_k \left[\sin^2\theta \left(k^2+m_1^2\right)^{1/2}
+\cos^2 \theta \left(k^2+m_2^2\right)^{1/2} \right]e^{ik(x-y)} \\
&G_3(x,y)\equiv V^{-1} \sum_k G_3(k) e^{ik(x-y)}
= V^{-1} \sum_k \left[\sin\theta \cos \theta\left[ \left(k^2+m_1^2\right)^{1/2}-\left (k^2+m_2^2\right)^{1/2}\right] \right]e^{ik(x-y)}
\end{align}
We can obtain the Renyi entropies $S_{3}^{(n)}$
and the entanglement entropy $S_{3}$ for the ground state 
 from 
(\ref{general r entropy}) and (\ref{general entropy}) as
\begin{align}
&\dfrac{S_3^{(n)} }{(2\pi)^{1-d}V }  \equiv \int d^{d-1}  \mathbf{k} s_3^{(n)}(k) =
 \int d^{d-1} \mathbf{k}  (1-n)^{-1} \left[n \ln (1-\xi (k))-\ln(1-\xi (k)^n ) \right]
,     \label{massive r entropy} \\
&\dfrac{S_3 }{(2\pi)^{1-d}V } \equiv \int d^{d-1}  \mathbf{k} s_3(k) =
\int d^{d-1} \mathbf{k}  \left[- \ln (1-\xi(k))-\dfrac{\xi (k)}{1-\xi(k)} \ln \xi (k) \right]
 ,\label{massive entropy}
\end{align}
where we have taken $V\rightarrow \infty$ and
$\sum_{k}\rightarrow V(2\pi)^{1-d}\int d^{d-1}  \mathbf{k}$ and
\begin{equation}
\xi(k) =\dfrac{1}{z(k)} \left[1-\sqrt{1-z(k)^2}\right] . \label{www}
\end{equation}
and
\begin{equation}
z(k)=\left[2G_1(k)G_2(k)G_3(k)^{-2} -1\right]^{-1}
\end{equation}
When $m_1=m_2$, $z(k)=0$ and $S_3=0$.
This is because $\phi$ and $\psi$ do not mix with each other when $m_1=m_2$.
For large $k$, we obtain
\begin{equation}
z(k) \approx \dfrac{1}{8k^4} \sin^2 \theta \cos^2 \theta (m_1^2-m_2^2)^2. \label{approx z}
\end{equation}
From (\ref{massive entropy}), (\ref{www}) and (\ref{approx z}), $S_3$ has UV divergence
for $d\ge 5$ and we obtain the leading  UV divergent term
\begin{align}
\dfrac{S_3 }{(2\pi)^{1-d}V } &=\dfrac{1}{4} \sin^2 \theta \cos^2 \theta (m_1^2-m_2^2)^2  \int d\Omega_{d-1} \int^\Lambda dk k^{d-6} \ln k  \\
&=\dfrac{1}{4} \sin^2 \theta \cos^2 \theta (m_1^2-m_2^2)^2  \int d\Omega_{d-1} \times
\begin{cases}
\dfrac{1}{2} (\ln \Lambda )^2  & for ~~~ d=5  \\
\dfrac{1}{d-5} \Lambda^{d-5} \ln \Lambda & for ~~~ d \geq 6 .
\end{cases}
\end{align}
Notice that the coefficient for $d=5$ is universal, i.e. it does not change by rescaling $\Lambda$.
For $d \leq 4$, $S_3$ is finite. Similarly for the Renyi entropy, we can find that $S^{(n)}_3$ gets divergent as $\Lambda^{d-5}$ for $d\geq 6$ and as $\log \Lambda $ for $d=5$, while it is finite for
$d\leq 4$.

Let us compute $S_3$ for $d \leq 4$ and for $|\theta| \ll1$.
In this case we obtain
\begin{equation}
z(k) \approx \dfrac{\theta^2}{2} \left( \sqrt{ \dfrac{k^2+m_1^2}{k^2+m_2^2} }
+ \sqrt{ \dfrac{k^2+m_2^2}{k^2+m_1^2} } -2  \right) .
\end{equation}
Thus we obtain
\begin{equation}
 \dfrac{S_3 }{(2\pi)^{1-d}V } \approx
-\dfrac{\theta^2}{4} \ln (\theta^2) \int d\Omega_{d-1} \int^\infty_0 dk k^{d-2}
 \left( \sqrt{ \dfrac{k^2+m_1^2}{k^2+m_2^2} }
+ \sqrt{ \dfrac{k^2+m_2^2}{k^2+m_1^2} } -2  \right) .
\end{equation}
We can perform the $k$ integral explicitly for $d=3$ which leads to
\begin{equation}
 \left. \dfrac{S_3 }{(2\pi)^{1-d}V } \right|_{d=3} \approx
-\dfrac{\theta^2}{4}  \left( \ln (\theta^2) \right) \pi (m_1-m_2)^2 .
\end{equation}

\section{Entanglement Entropy Calculations using Boundary States}

In this section, we present another useful method for the calculation of entanglement entropy between two CFTs. This is based on the Euclidean replica method and the folding method  as we will explain later. We will study the massless interaction model (\ref{conformal action}) and massive interaction model (\ref{massive action}) in $d=2$ dimensions, introduced in the previous section. Refer to \cite{FK} for a similar replica method calculations at finite temperature for Tomonaga-Luttinger liquids.

Since we are going to use an operator representation based on mode expansions for practical calculations, let us start with our field theory convention. Let us focus on the field theory defined by the massless interaction model (\ref{conformal action}) in two dimensional spacetime, whose coordinate is denoted by $(t,x)$.
We compactify the space coordinate $x$ on a circle so that it has the periodicity $x\sim x+2\pi$. It is straightforward to change the radius of this circle in the analysis below.

In the Hamiltonian description, we define the momentum operators
\ba
&& \pi_{\phi}=\de_{t}\phi+\f{\lambda}{2}\de_t \psi,\no
&& \pi_{\psi}=\de_t \psi+\f{\lambda}{2}\de_t \phi.
\ea
This leads to the Hamiltonian
\be
H=\f{1}{8\pi}\int dx \left[\f{1}{1-\f{\lambda^2}{4}}(\pi_\phi^2+\pi_{\psi}^2-\lambda \pi_{\phi}\pi_{\psi})+(\de_x\phi)^2+(\de_x\psi)^2+\lambda (\de_x\phi)(\de_x\psi)
\right] ,
\ee
where we changed the overall factor of the Hamiltonian compared with (\ref{conformal action}) so that the expression of quantized fields looks simpler (this corresponds to
$\al=2$ convention in the string world-sheet theory).

Therefore we can express
\be
H=H_{0}+H_{int},
\ee
where
\ba
H_0&=& \f{1}{8\pi}\int dx\left[\pi_{\phi}^2+\pi_{\psi}^2
+(\de_x\phi)^2+(\de_x\psi)^2\right], \no
H_{int}&=& \f{1}{8\pi}\int dx  \left[\f{1}{1-\lambda^2/4}\left(\f{\lambda^2}{4}\pi^2_{\phi}+\f{\lambda^2}{4}\pi^2_{\psi}
-\lambda \pi_{\phi}\pi_{\psi}\right)+\lambda(\de_x \phi)(\de_x\psi)\right]. \label{int}
\ea

We can quantize the free theory described by the Hamiltonian $H_0$ as usual. The mode expansions of scalar fields look like
\ba
 && \phi(t,x)=i\sum_{n}\f{1}{n}\left(\ap_n e^{-in(t-x)}+\ti{\ap}_n e^{-in(t+x)}\right),\no
 && \psi(t,x)=i\sum_{n}\f{1}{n}\left(\beta_n e^{-in(t-x)}+\ti{\beta}_n e^{-in(t+x)}\right),
 \label{mode}
\ea
where the oscillators $\ap_n$ and $\beta_n$ satisfy the commutation relations:
\be
[\ap_n,\ap_m]=[\beta_n,\beta_m]=[\ti{\ap}_n,\ti{\ap}_m]=[\ti{\beta}_n,\ti{\beta}_m]=n\delta_{n+m,0}.
\ee
We omitted the zero modes $n=0$ since they do not contribute in our calculations as we will mention later.

 When we study the time evolution of the full Hamiltonian $H=H_0+H_{int}$, then we can use the interaction picture:
\be
e^{-iHt}=e^{-iH_0 t}\cdot {\mathcal P}\exp\left[-i\int^{t}_0 ds H_{int}(s)\right],
\ee
where we defined
\be
H_{int}(s)=e^{iH_0 s}\cdot H_{int}\cdot  e^{-iH_0 s}.
 \ee
 Here ${\mathcal P}$ is the path-ordering such
 that the late time operator is placed in the left.

In this interaction picture we can regard $\pi_{\phi}$ and $\pi_{\psi}$ in $H_{int}$ as
$\dot{\phi}(s,x)$ and $\dot{\psi}(s,x)$ defined by the free field ones (\ref{mode}).
Thus, the interaction (\ref{int}) is written as follows (up to a certain numerical constant)
\be
H_{int}(s)=-\f{\lambda}{2} \sum_{n=-\infty}^{\infty} e^{-2ins}[\ap_n\ti{\beta}_{n}+\beta_n\ti{\ap}_{n}]+O(\lambda^2),
\label{inta}
\ee
where we only make explicit the leading term in the weak coupling limit $\lambda\to 0$.

\subsection{Replica Calculation}

We would like to employ the replica method to calculate the entanglement ($n$-th Renyi) entropy between the two scalar field theories $\phi$ and $\psi$: $S^{(n)}_{ent}=\f{1}{1-n}\log\mbox{Tr}(\rho_{\psi})^n$, where $\rho_{\psi}=\mbox{Tr}_{\phi}\rho_{tot}$ is the reduced density matrix defined by tracing out the Hilbert space for the field $\phi$. The entanglement entropy is obtained by taking the limit $n\to 1$. The replica method calculation for quantum field theories when we geometrically define the subsystem is well-known (see e.g. \cite{CC}) and we consider a straightforward modification of this in our case.

In the path integral description, we take the Euclidean time to be $-\infty<\tau<\infty$ and the one dimensional space to be $0\leq x \leq 2\pi$. The ground state wave functional $\Psi_0[\phi,\psi]$ at $\tau=0$ is given by path-integrating the scalar fields from $\tau=-\infty$ to $\tau=0$. The total density matrix is given by
\be
\rho_{tot}[(\phi_1,\psi_1),(\phi_2,\psi_2)]=
\Psi^*_0[\phi_1,\psi_1]\cdot \Psi_0[\phi_2,\psi_2].
\ee
The reduced density matrix is obtained by path-integrating over the field $\phi$:
\be
\rho_{\psi}[\psi_1,\psi_2]=\int [D\phi] \Psi^*_0[\phi,\psi_1]\cdot \Psi_0[\phi,\psi_2].
\ee

\begin{figure}[ttt]
   \begin{center}
     \includegraphics[height=5cm]{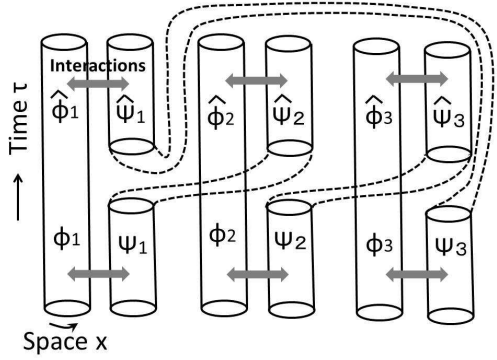}
   \end{center}
   \caption{The replica calculation for the trace $\mbox{Tr}(\rho_{\psi})^n$. We depicted the picture assuming $n=3$.}\label{fig:replica}
\end{figure}

To calculate $\mbox{Tr}(\rho_{\psi})^n$, let us consider $n$ copies of scalar fields (replicas):
\be
(\phi_i(\tau,x),\psi_i(\tau,x)), \ \ i=1,2,\ddd,n.
\ee
Then the trace $\mbox{Tr}(\rho_{\psi})^n$ is given the partition function of this system with $2n$ scalar fields with the following boundary condition at $\tau=0$:
\be
\phi_{i}(\delta,x)=\phi_{i}(-\delta,x),\ \ \ \psi_{i+1}(\delta,x)=\psi_{i}(-\delta,x),
\label{bcd}
\ee
where $\delta>0$ takes an infinitesimally small value (see Fig.\ref{fig:replica}).
If we want to generalize the above replica formulation to more general cases of excited states, we just need to replace the ground state wave functional $\Psi_0$ with that for an excited state.

One way to deal with this boundary condition is to use the folding trick (see e.g. \cite{Bac}). We regard the fields $(\phi_i(\tau,x),\psi_i(\tau,x))$ for $\tau>0$ as another fields $(\hat{\phi}_i(-\tau,x),\hat{\psi}_i(-\tau,x))$ as in Fig.\ref{fig:replica}. We can define the mode expansions for these fields in a similar way
\ba
&& \hat{\phi}_i(t,x)=i\sum_{n}\f{1}{n}\left(\hat{\ap}^{(i)}_n e^{-in(t-x)}+\hat{\ti{\ap}}^{(i)}_n e^{-in(t+x)}\right),\no
 && \hat{\psi}_i(t,x)=i\sum_{n}\f{1}{n}\left(\hat{\beta}^{(i)}_n e^{-in(t-x)}+\hat{\ti{\beta}}^{(i)}_n e^{-in(t+x)}\right).
\ea
We also define their interacting and non-interacting Hamiltonian to be $\hat{H}(\equiv
\hat{H}_0+\hat{H}_{int})$ and $\hat{H}_0$, respectively. They are simply given by putting the hat: $a\to \hat{a}$ for all oscillators in $H$ and $H_0$.

In the presence of such doubled degrees of freedom, we can equivalently compute $\mbox{Tr}(\rho_{\psi})^n$ as a partition function on the half space defined by
$-\infty<\tau\leq 0$ and $0\leq x\leq 2\pi$, which is a particular example of the so called
folding procedure. Then we can rewrite (\ref{bcd}) into (by taking
$\delta\to 0$ limit):
\ba
&& \phi_i(0,x)-\hat{\phi}_i(0,x)=0,\ \ \ \de_{\tau}(\phi_i(0,x)+\hat{\phi}_i(0,x))=0, \no
&& \psi_i(0,x)-\hat{\psi}_{i+1}(0,x)=0,\ \ \ \de_{\tau}(\psi_i(0,x)+\hat{\psi}_{i+1}(0,x))=0.
\ea

In the quantized theory, we can describe such boundary conditions in terms of boundary states
$|B_{(n)}\lb$ (see e.g. \cite{RSB,DL} for reviews), which are defined by
\ba
&& (\phi_i(0,x)-\hat{\phi}_i(0,x))|B_{(n)}\lb=0,\ \ \ \de_{\tau}(\phi_i(0,x)+\hat{\phi}_i(0,x))|B_{(n)}\lb=0, \no
&& (\psi_i(0,x)-\hat{\psi}_{i+1}(0,x))|B_{(n)}\lb=0,\ \ \ \de_{\tau}(\psi_i(0,x)+\hat{\psi}_{i+1}(0,x))|B_{(n)}\lb=0. \label{bdfd}
\ea

This is solved in terms of oscillators as follows:
\be
|B_{(n)}\lb ={\mathcal N_{(n)}}\cdot \exp\left(-\sum_{i=1}^n\sum^\infty_{m=1}\f{1}{m}
(\ti{\ap}^{(i)}_{-m}\hat{\ap}^{(i)}_{-m}+\ap^{(i)}_{-m}\hat{\ti{\ap}}^{(i)}_{-m}
+\ti{\beta}^{(i)}_{-m}\hat{\beta}^{(i+1)}_{-m}+\beta^{(i)}_{-m}\hat{\ti{\beta}}^{(i+1)}_{-m})
\right)|0_{(n)}\lb, \label{bsta}
\ee
where the vacuum $|0_{(n)}\lb$ for the $n$-replicated theory is defined by
requiring that it is annihilated by any oscillator with a positive mode number as usual. The above conditions are equivalent to
\ba
(\ap^{(i)}_{-m}+\hat{\ti{\ap}}^{(i)}_m)|B_{(n)}\lb
\!=\!(\ti{\ap}^{(i)}_{-m}+\hat{\ap}^{(i)}_m)|B_{(n)}\lb\!=\!
(\beta^{(i)}_{-m}+\hat{\ti{\beta}}^{(i+1)}_m)|B_{(n)}\lb
\!=\!(\ti{\beta}^{(i)}_{-m}+\hat{\beta}^{(i+1)}_m)|B_{(n)}\lb\!=\!0.
\ea

The coefficient ${\mathcal N_{(n)}}$ in (\ref{bsta}) represents
 the normalization of the boundary state. We can determine ${\mathcal N_{(n)}}$  from the open-closed duality of the cylinder amplitudes \cite{Cardy,RSB,DL}. However, it is clear that the calculation is just the $n$-th power that of the standard two free scalars without any replicas. Therefore we can conclude
\be
{\mathcal N_{(n)}}={\mathcal N}^n,
\ee
where ${\mathcal N}(={\mathcal N_{(1)}})$ is the standard normalization constant for free two scalar field theories. Having defined the boundary state at $\tau=0$ or equally $t=0$, we will study the evolution under the Lorentzian time $t$ below.

Now, consider a pure state $|\Phi\lb(=|\Phi_{(1)}\lb)$, which is a general excited state. We can define a corresponding excited state
$|\Phi_{(n)}\lb$ in the $n$ replicated theory. Then the Renyi entanglement entropy between $\phi$ and $\psi$ for the reduced density matrix $\rho_\psi=\mbox{Tr}_{\phi}|\Phi\lb \la \Phi|$, can be computed as
\be
S^{(n)}_{ent}=\f{1}{1-n}\log \mbox{Tr}\rho_{\psi}^n=\f{1}{1-n}\log\left[ \f{\la \Phi_{(n)}|B_{(n)}\lb}{\left(\la \Phi_{(1)}|B_{(1)}\lb\right)^n}\right].
\ee
In the previous calculations, we neglected the zero modes of the fields $\phi$ and $\psi$. Indeed, we can easily confirm that there is no contribution from zero modes as the contributions cancel completely
in the ratio $\la \Phi_{(n)}|B_{(n)}\lb / \left(\la \Phi_{(1)}|B_{(1)}\lb\right)^n$.

If we consider the time evolution under the free Hamiltonian $H_0$, it will be described by
the unitary operator $e^{-i(H_0-\hat{H}_0)t}$. This is because the time  flows in the opposite direction for $\hat{\phi}$ and $\hat{\psi}$ as we folded the time direction. Therefore time evolutions under the free Hamiltonian is trivial as follows
\be
e^{-i(H_0-\hat{H}_0) t}|B_{(n)}\lb=|B_{(n)}\lb,
\ee
which does not change the entanglement entropy.
This fact is easily understood because in this case the time evolution simply leads to the unitary transformation of $\rho_A$:  $\rho_A\to e^{-iH_0t}\rho_A e^{iH_0 t}$ and thus the entropy $S^{(n)}_A$ does not change.

For example, we can consider the following entangled state as an example of excited state
\be
|\Phi_{(n)}\lb=(2m)^{-n}\prod_{i=1}^n(\ap^{(i)}_{-m}+\beta^{(i)}_{-m})
(\hat{\ti{\ap}}^{(i)}_{-m}+\hat{\ti{\beta}}^{(i)}_{-m})|0_{(n)}\lb.
\ee
Notice that any state $|\Phi_{(n)}\lb$ should have the replica symmetry which exchanges
$(\phi_i,\psi_i)$ with $(\phi_j,\psi_j)$. Moreover, at $t=0$, $|\Phi_{(n)}\lb$ should have an
inversion symmetry which exchanges $(\ap_{n},\ti{\ap}_n,\beta_n,\ti{\beta}_n)$ with $(\hat{\ti{\ap}}_n,\hat{\ap}_n,\hat{\ti{\beta}}_n,\hat{\beta}_n)$, where the chirality is also exchanged owing to the boundary condition (\ref{bdfd}).

We find
\ba
\la \Phi_{(n)}|B_{(n)}\lb &=&\la 0_{(n)}|\prod_{i=1}^n(\ap^{(i)}_{m}+\beta^{(i)}_{m})
(\hat{\ti{\ap}}^{(i)}_{m}+\hat{\ti{\beta}}^{(i)}_{m})|B\lb \no
&=&(-1)^n\cdot \la 0_{(n)}|\prod_{i=1}^n(\ap^{(i)}_{m}+\beta^{(i)}_{m})
(\ap^{(i)}_{-m}+\beta^{(i-1)}_{-m})|B\lb \no
&=& 2\cdot (-1)^n\cdot  \left(\f{{\mathcal N}}{2}\right)^n.
\ea
Thus we simply get
\be
S^{(n)}_A=\f{\log 2^{1-n}}{1-n}=\log 2.
\ee
This is naturally understood as in \cite{NNT} because the state we are looking at is
\be
|\Psi\lb=\f{1}{2m}(\ap_{-m}+\beta_{-m})|0\lb,
\ee
in the original (unreplicated) theory and this is an maximally entangled state (EPR state).

\subsection{Time-dependent Formulation}

In order to get a non-trivial time-dependence we need to add the interaction between $\phi$ and
$\psi$. We would like to study the case where the interaction is given by  (\ref{inta}).
For this purpose let us study the following state
\be
|\Phi_n(t)\lb=e^{-i(H-\hat{H})t}|0_{(n)}\lb.
\ee
Note that $|0_{(n)}\lb$ is the vacuum for the free Hamiltonian $H_0-\hat{H}_0$, while the time evolution by $H-\hat{H}$ is non-trivial.

Below we would like to compute the Renyi entanglement entropy up to $O(\lambda^2)$ in the weak coupling expansions. Let us work with the interaction picture. In our replicated model with the folding, the interaction Hamiltonian looks like
\ba
&& H_{int}(s)-\hat{H}_{int}(s) \no
&& =-\f{\lambda}{2} \sum_{i=1}^n\sum_{m=-\infty}^{\infty} \left[ e^{-2ims}(\ap^{(i)}_m\ti{\beta}^{(i)}_{m}+\beta^{(i)}_m\ti{\ap}^{(i)}_{m})
-e^{2ims}(\hat{\ap}^{(i)}_m\hat{\ti{\beta}}^{(i)}_{m}
+\hat{\beta}^{(i)}_m\hat{\ti{\ap}}^{(i)}_{m})\right]
+O(\lambda^2),
\label{intaa}
\ea

It is easy to see that there is no $O(\lambda)$ contributions to $S^{(n)}_{ent}$ and the leading non-trivial order is $O(\lambda^2)$. Therefore we obtain the following behavior
\ba
&& \la \Phi_{(n)}(t)|B_{(n)}\lb \no
&& =\la 0_{(n)}| {\mathcal P}\exp\left[i\int^{t}_0 ds (H_{int}(s)-\hat{H}_{int}(s))\right]|B_{(n)}\lb \no
&&={\mathcal{N}}^n\cdot \left(1+G_{n}\lambda^2+O(\lambda^3)\right).
\ea
The Renyi entropy can be estimated as
\ba
&& S^{(n)}_{ent} =\f{1}{1-n}\log \mbox{Tr}(\rho_A)^n \no
&& =\f{1}{1-n}\log\left(1+(G_{n}-nG_{1})\lambda^2+O(\lambda^3)\right) \no
&& =\f{\lambda^2}{1-n}(G_{n}-nG_{1})+O(\lambda^3). \label{renyibs}
\ea

Moreover, we can  confirm that $O(\lambda^2)$ contributions only come from the square of the above $O(\lambda)$ term in
$H_{int}(s)-\hat{H}_{int}(s)$ via the Taylor expansion of
\be
e^{-i(H-\hat{H})t}=e^{-i(H_0-\hat{H}_0)t}\cdot {\mathcal P}\exp\left[-i\int^{t}_0 ds (H_{int}(s)-\hat{H}_{int}(s))\right].
\ee
Thus the coefficient $g_{n}$ is the sum of three contributions: $(H_{int})^2$, $(\hat{H}_{int})^2$
and $H_{int}\cdot \hat{H}_{int}$. It is easy to see that the former two contributions behave
$G_{n}\propto n$ for any positive integer $n$. Thus these do not contribute to the Renyi
entropy (\ref{renyibs}). Thus the only possibility is that from $H_{int}\cdot \hat{H}_{int}$, whose contribution to $G_n$ is denoted by $g_n$.
This is calculated as follows
\ba
&& {\mathcal{N}}^n\cdot g_{n}=\f{1}{4}\int^t_0 ds_1 \int^t_0 ds_2 \sum_{m=1}^{\infty}e^{2im(s_2-s_1)}\sum_{i.j=1}^n
\la 0_{(n)}|(\ap^{(i)}_m\ti{\beta}^{(i)}_{m}+\beta^{(i)}_m\ti{\ap}^{(i)}_{m})
(\hat{\ap}^{(j)}_m\hat{\ti{\beta}}^{(j)}_{m}
+\hat{\beta}^{(j)}_m\hat{\ti{\ap}}^{(j)}_{m})|B_{(n)}\lb \no
&& =\f{1}{4}\int^t_0 ds_1 \int^t_0 ds_2 \sum_{m=1}^{\infty}e^{2im(s_2-s_1)}\sum_{i.j=1}^n
\la 0_{(n)}|(\ap^{(i)}_m\ti{\beta}^{(i)}_{m}+\beta^{(i)}_m\ti{\ap}^{(i)}_{m})
(\ti{\ap}^{(j)}_{-m}\beta^{(j)}_{-m}
+\ti{\beta}^{(j)}_{-m}\ap^{(j)}_{-m})|B_{(n)}\lb.\nonumber \\ \label{intbc}
\ea
Then it is easy to see that $g_{n}$ is vanishing except for $n=1$. In our perturbation theory, we find
\be
S^{(n)}_{ent}=\f{n}{n-1}g_1 \lambda^2+O(\lambda^3). \label{fmee}
\ee
Note that this perturbation theory is valid except for the entanglement entropy limit $n\to 1$.
This is because in general we expect a term in $S^{(n)}_{ent}$ which is proportional to $\lambda^{2n}$ originally corresponding to the higher order term and this becomes $O(\lambda^2)$ in the $n\to 1$ limit.

\subsection{Explicit Evaluations for Massless Interaction}

Now we can explicitly evaluate $g_1$ from (\ref{intbc}) as follows
\ba
g_1&=&\int^t_0 ds_1 \int^t_0 ds_2 \sum_{m=1}^{\infty}\f{m^2}{2}\cdot e^{2im(s_2-s_1)}  \no
&=& \f{1}{2}\sum_{m=1}^\infty \sin^2(mt). \label{gone}
\ea
Note that this expression indeed agrees with our previous result (\ref{timentropy}) obtained from the real time formulation. Since this is clearly UV divergent, let us introduce a UV regular $\ep$ simply by adding the weight $e^{-m\ep}$ in the sum (\ref{gone}). This effectively removes the interactions for high energy modes $m>1/\ep$.  In a realistic experiment, indeed we cannot turn on the interaction for such a high energy mode.

\begin{figure}[ttt]
   \begin{center}
     \includegraphics[height=4cm]{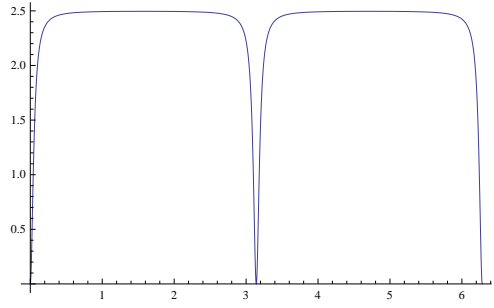}
   \end{center}
   \caption{A plot of $g_1=\f{n-1}{n\lambda^2}S^{(n)}_{ent}$ as a function of $t$ in the case of massless interaction. We chose $\ep=0.1$.}\label{fig:PL}
\end{figure}

This regularization leads to
\be
g_1=\f{1}{2}\sum_{m=1}^\infty  \sin^2(mt) e^{-m\ep}=\f{1}{4(e^{\ep}-1)}
-\f{e^{\ep}\cos(2t)-1}{4(e^{2\ep}-2e^{\ep}\cos(2t)+1)},
\ee
which is plotted in Fig.\ref{fig:PL}. At early time it increase like $g_1\propto t^2$ and at $t=\pi/2$, it reaches the maximum value $g_1(\pi/2)=\f{e^{\ep}}{2(e^{2\ep}-1)}$. It has the periodicity $g_1(t+\pi)=g_1(t)$ and this is peculiar to the free field theories on a circle.
If we want to recover the result in terms of momenta and
the general periodicity $L$ of the compact direction, we just need to replace $\sum_{m}$ with
$(L/2\pi)\int dk$ and $mt$ with $kt$.

When $\ep$ is infinitesimally small, we find that the system almost instantaneously reaches the maximal entanglement entropy within the time of order $\ep$. The maximum value behaves like
$g_1(\pi/2)\simeq \f{1}{4\ep}+O(\ep)$. Note that this instantaneous increasing of the entanglement entropy occurs because the interaction introduces the entanglement between the two scalar field theories homogeneously at the same time. This is in contrast with evolutions of entanglement entropy under quantum quenches \cite{caca} or local operator excitations \cite{NNT}, where the causal propagation of entangled pairs play an important role.

It is also useful to analyze the evolution in the Euclidean time $\tau$ evolution instead of the real time $t$. This is realized by replacing the time integral of interactions as follows
\ba
&& P\exp\left[-i\int^{t}_0 ds (H_{int}(s)-\hat{H}_{int}(s))\right] \no
&& \to P\exp\left[-i\int^{-i\tau}_0 ds_1 H_{int}(s_1)+i\int^{i\tau}_{0} ds_2 \hat{H}_{int}(s_2)\right].
\ea
This leads to the evaluation
\ba
g_1&=&\int^{-i\tau}_0 ds_1 \int^{i\tau}_0 ds_2
\sum_{m=1}^{\infty}\f{m^2}{2}e^{2im(s_2-s_1)-2m\ep} \no
&=& \f{1}{8}\sum_{m=1}^\infty e^{-m\ep}(1-e^{-2m\tau})^2. \label{goned}
\ea

Assuming $\ep$ is infinitesimally small, we find
\be
g_1=\f{1}{8\ep}-\f{e^{4\tau}+4e^{2\tau}+1}{16(e^{4\tau}-1)}+O(\ep).
\ee
The ground state for $H=H_{0}+H_{int}$ corresponds to the
limit $\tau=\infty$, which reads $g_1|_{\tau=\infty}=\f{1}{8\ep}-\f{1}{16}$. This agrees with the previous result (\ref{entprl}) and (\ref{entml}) based on the real time formalism.

\subsection{Results for Massive Interaction}
In the case where the interaction is massive, we consider the following interaction
\be
H_{int}=\f{1}{8\pi}\int dx (A\phi^2+B\psi^2+2C\phi\psi), \label{massiveiy}
\ee
with the same $H_0$ as before. In this case we will not have any UV divergence and have a smooth time dependence even without the UV cut off as we will see.
\begin{align*}
H_{int}(s)=\f{1}{4}\sum_{m=-\infty}^{+\infty}\f{1}{m^2}&
\Bigg\{\left[A\left(\alpha_{m}\alpha_{-m}
+\tilde{\alpha}_{m}\tilde{\alpha}_{-m}\right)
+B\left(\beta_{m}\beta_{-m}+\tilde{\beta}_{m}\tilde{\beta}_{-m}\right)\right]
+2C\left(\alpha_{m}\beta_{-m}
+\tilde{\alpha}_{m}\tilde{\beta}_{-m}\right)\\&
-2e^{-2ims}\left[A\alpha_{m}\tilde{\alpha}_{m}
+B\beta_{m}\tilde{\beta}_{m}+C\left(\alpha_{m}\tilde{\beta}_{m}
+\beta_{m}\tilde{\alpha}_{m}\right)\right]\Bigg\}
\end{align*}

\begin{figure}
\centering
\includegraphics[scale=.8]{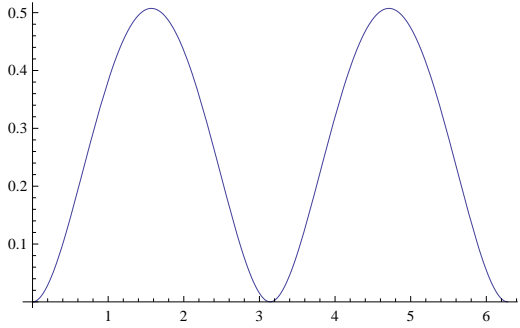}
\caption{A plot of $\f{n-1}{nC^2}S^{(n)}_{ent}$ in the massive interaction case.}
\label{fig:tht}
\end{figure}

We can proceed the calculations of Renyi entanglement entropy up to the quadratic order of the interaction (\ref{massiveiy}) using our boundary state (\ref{bsta}) in a way very similar to (\ref{intbc}). We find that terms proportional to $A^2$ and $B^2$ only lead to results which behave $G_n\propto n$ and thus they do not contribute the Renyi entropy as explained in (\ref{renyibs}). The term proportional to $C^2$ becomes non-zero only for $n=1$ and thus contribute to the Renyi entropy. In the end we obtain the following result of $S^{(n)}_{ent}$:
\ba
S^{(n)}_{ent}(t)&=&\f{nC^2}{2(n-1)}\int^t_0 ds_1\int^t_0 ds_2 \sum_{m=1}^\infty\f{e^{2im(s_2-s_1)}}{m^2} \no
&=& \f{nC^2}{2(n-1)}\sum_{m=1}^\infty \f{\sin^2(mt)}{m^4} \no
&=& \f{n}{360(n-1)}(\pi^4-45\cdot\mbox{Li}_4(e^{-2it})-45\cdot\mbox{Li}_4(e^{2it})),
\ea
where Li$_4(x)$ is a Polylog function defined by Li$_4(x)=\sum_{k=1}^\infty \f{x^k}{k^4}$.
This Renyi entropy takes finite values even in the absent of the UV cut-off and it is plotted in Fig.\ref{fig:tht}.

The Euclidean time evolution can also be obtained as follows
\be
S^{(n)}_{ent}(\tau)
= \f{nC^2}{8(n-1)}\sum_{m=1}^\infty \f{(1-e^{-2m\tau})^2}{m^4}.
\ee
By taking $\tau\to \infty$ we get the result for the ground state
$S^{(n)}_{ent}(\infty)=\f{\pi^4nC^2}{720(n-1)}$.

 The reason why $S^{(n)}_{ent}$ for the massive interaction is free from UV divergences is that the massive interactions can be negligible in the high energy region. This is consistent with our previous results using the real time formalism, where it was shown that the Renyi entropy for the ground state in the massive interaction model is finite in $d\leq 4$ dimensions.

 Note also that as opposed to the massless interaction case, $S_{ent}$ grows slowly in the massive interaction case. If we take the non-compact limit $L\to \infty$ of the space direction (remember we set $L=2\pi$ in our calculations in this section), then we find
 the simple behavior $S_{ent}\propto C^2 t^2$.

\section{Generalized Holographic Entanglement Entropy}

Finally we would like to discuss a holographic counterpart of entanglement between two interacting CFTs. One may naturally come up with a holographic model where two (conformal) gauge theories, such as two ${\mathcal N=4}$ super Yang-Mills theories, are interacting with each other via the interaction of the form
\be
{\mathcal L}_{int}=g O_1O_2. \label{intg}
\ee
Here $g$ is a coupling constant, while $O_1$ and $O_2$ are single trace operators in
CFT$_1$ and CFT$_2$, respectively. If we assume that each of two gauge theories (CFTs)
has its gravity dual, then the interaction (\ref{intg}) can be regarded as a multi-trace
deformation \cite{WM} and thus this can
be taken into account as a deformation of boundary condition of supergravity fields in the bulk AdS \cite{Massive}.
In this case, it is obvious the entanglement entropy does not arise at the tree level but does at one loop order because the metric is modified only after we incorporate quantum corrections of supergravity. When the interaction (\ref{intg}) is marginal or relevant, then this holographic model
is expected to be similar to our massless interaction model or massive one defined in section 2,
respectively.

\begin{figure}
\begin{center}
\includegraphics[scale=.3]{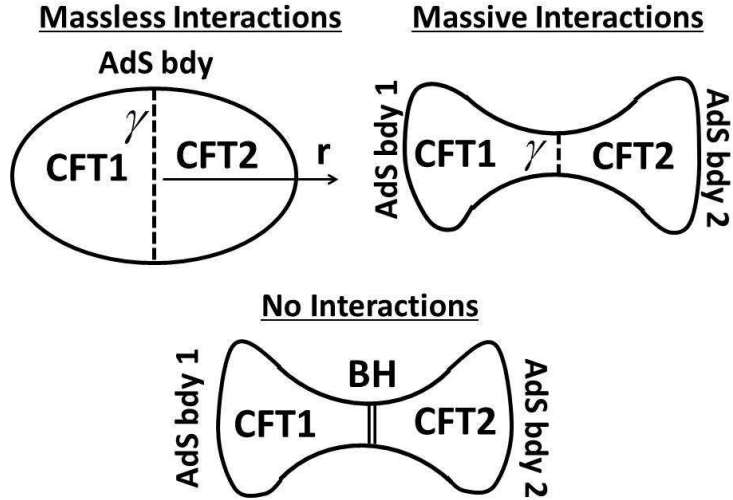}
\caption{Geometries of gravity duals for two CFTs (CFT$_1$ and CFT$_2$) with non-vanishing quantum entanglement between them.  The surface $\gamma$ denotes the minimal surface which computes the entanglement entropy between them. The upper left picture describes a geometry for two CFTs with  massless (or marginal) interactions between them. The upper right one expresses a possibility of geometry with massive interactions, which should be taken cautiously as we mention in section \ref{fink}. In this geometry for the massive interactions we need to further identify the two AdS boundaries. These setups should be distinguished from the lower picture, where the two CFTs are entangled without any interactions, such as in the thermofield double construction dual to a AdS black hole.}
\label{fig:geo}
\end{center}
\end{figure}

Even though such one loop calculations of entanglement entropy are an interesting future problem, here we will view this problem for a massless interaction model from a slightly different angle so that entanglement entropy can be estimated classically by generalizing the minimal surface formula of holographic entanglement entropy \cite{RT}. We will argue that in such a case we need to take a minimal surface which divides the internal manifold (e.g. S$^5$) into two halves as opposed to the standard prescription \cite{RT}, where a minimal surface divides a time slice in AdS into two halves. This analysis should be distinguished from a system of two entangling CFTs without interactions, where its gravity dual is given by an AdS black hole geometry \cite{EBH}. In our case, on the other hand, the two CFTs are interacting directly and thus they can communicate causally in their gravity dual. We summarized a sketch of geometries of gravity duals for two entangling CFTs in Fig.\ref{fig:geo}, whose details will be explained later.

\subsection{Multiple D3-branes in Supergravity}

Let us consider $N$ D3-branes which are parallel and are separated from each other generically.
This corresponds to a Coulomb branch of the four dimensional $U(N)$ ${\cal N}=4$ super Yang-Mills theory. We express its six transverse scalars by $\vec{\Phi}_{ab}=(\Phi^1_{ab},\Phi^2_{ab},\ddd,\Phi^6_{ab})$, where $a,b=1,2,\ddd ,N$.
In any vacua of Coulomb branch, the $N\times N$ matrices $\vec{\Phi}$ can be diagonalized at the same time and their eigenvalues are denoted by
$\vec{\vp}_a=(\vp^1_{a},\vp^2_a,\ddd , \vp^6_a)$. The type IIB supergravity solution in the near horizon limit corresponds to this
Coulomb branch \cite{Maldacena,Kraus} is given by
\be
ds^2=H^{-1/2}(\vec{y}) dx^\mu dx_\mu +H^{1/2}(\vec{y})dy^i dy^i,
\ee
where $\mu=0,1,2,3$ and $i=1,2,\ddd, 6$. We defined
\be
H(\vec{y})=\sum_{a=1}^{N}\f{4\pi\al^2g_s}{|\vec{y}-\vec{y}_a|^4},
\ee
where $\vec{y}_a\equiv 2\pi\al\vec{\vp}_a$ are positions of each D3-brane.

If D3-branes are situated at a single point, this becomes the AdS$_5\times$ S$^5$ geometry
(we set $|y|=r$):
\be
ds^2=\f{r^2}{R^2}(dx^\mu dx_\mu)+\f{R^2}{r^2}(dr^2+r^2 d\theta^2+r^2\sin^2\theta d\Omega_4^2),
\label{ads}
\ee
where $R^4=4\pi\al^2 Ng_s$. Here $\theta$ takes the values $0\leq \theta\leq \pi$ and $\Omega_4$
denotes the coordinate of S$^4$ with the unit radius. It is useful for a later purpose to note
that the volumes of $S^4$ and $S^5$ with unit radius are given by Vol(S$^4$)$=\f{8}{3}\pi^2$ and Vol(S$^5$)$=\pi^3$. The ten dimensional Newton constant is $G^{(10)}_N=8\pi^6\al^4g_s^2$, which leads to
\be
\f{R^8}{G^{(10)}_N}=\f{2N^2}{\pi^4}.
\ee

\subsection{$SU(N/2)\times SU(N/2)$ Solution and Entanglement}

Now we can consider the case where $N/2$ D3-branes are placed at
$\vec{y}=(-l,0,0,0,0,0)$ and the other $N/2$ D3-branes are at $\vec{y}=(l,0,0,0,0,0)$.
In this case, the function $H$ is given by
\be
H=\f{R^4}{2(r^2+l^2-2rl\cos\theta)^2}+\f{R^4}{2(r^2+l^2+2rl\cos\theta)^2}.
\label{col}
\ee

The dual gauge theory has the unbroken gauge group $SU(N/2)\times SU(N/2)$ and fields which are bi-fundamental with respect to each of the two $SU(N/2)$ groups are massive with the mass
\be
M=\f{l}{\pi\al},
\ee
as can be easily understood by estimating the mass of open string. If we focus on the physics below this mass scale $M$ and take the large $N$ limit, we can reliably integrate out the bi-fundamental massive modes and can regard that the system is described by two $SU(N/2)$ gauge theories (CFTs) interacting with each other. We call them CFT$_1$ and CFT$_2$. In this setup, the strength of the interaction between them is parameterized by
$1/M^2$, which estimates the propagator of massive modes which include the interactions.
Motivated by this we would like to parameterize the strength of interactions between
CFT$_1$ and CFT$_2$ by the dimensionless coupling:
\be
g=\f{\Lambda^2}{M^2}, \label{cpc}
\ee
where $\Lambda$ is the UV cut off of this theory.
In this way, this setup looks similar to the ones we discussed in the previous sections, though here we are considering strongly coupled gauge theories. This should qualitatively correspond to the massless interaction model as the interaction exist at any energy scale, while in the massive interaction model it does only in the low energy $E<<\Lambda$.

We would like to argue that the entanglement entropy $S_{ent}$ between these two CFTs are given by the holographic entanglement formula
\be
S_{ent}=\f{\mbox{Area}(\gamma)}{4G_N}, \label{hee}
\ee
by choosing $\gamma$ to be the area of minimal surface which separates the two groups of $N/2$ D3-branes (see the upper left picture in Fig.\ref{fig:geo}). From a symmetrical reason, it is clear that $\gamma$ is given by the eight dimensional surface
\be
\gamma: t=0,\ \ \ \theta=\f{\pi}{2}. \label{ents}
\ee

The corresponding entanglement entropy (\ref{hee}) is computed as
\ba
S_{ent}&=&\f{V_3R^2\mbox{Vol}(\mbox{S}^4)}{4G^{(10)}_N}\int^{r_{UV}}_{0} \f{r^4}{r^2+l^2} dr  \no
&=& \f{4}{3\pi^2}\cdot\f{N^2V_3}{R^6} \int^{r_{UV}}_{0} \f{r^4}{r^2+l^2} dr  , \label{enttwodp}
\ea
where $V_3$ is the volume in the $(x^1,x^2,x^3)$ direction; $r_{UV}$ is the cut off in the radial direction and is related to the UV cut off $\Lambda$ in the CFT via the usual UV-IR relation \cite{SWi}:
\be
r_{UV}=\Lambda R^2.
\ee

 In order to justify our interpretation of the D3-brane system as two interacting CFTs, we need to require
\be
r_{UV}\ll l.  \label{cond}
\ee
In this case we can approximate (\ref{enttwodp}) by
\ba
S_{ent}&\simeq & \f{4}{15\pi^2}\cdot\f{N^2V_3r_{UV}^5}{R^6l^2}
=\f{16N^2V_3}{15\pi^3}\lambda g\Lambda^3,
 \label{enttwod}
\ea
where $\lambda=Ng_s$ is the 't Hooft coupling of the gauge theory and $g$ is the effective coupling between the two CFTs defined in  (\ref{cpc}). It is clear from this expression that the entanglement entropy vanishes if $g=0$ as expected from the gauge theory.
 If we interpolate our result to the boarder region $r_{UV}\sim l$, where our argument can qualitatively be applied, we find the behavior
$S_{ent}\sim N^2V_3 \lambda \Lambda^3$. In this way, our holographic analysis reproduced the volume law which we found in the scalar field theory calculations for the massless interaction (\ref{entprkl}).

\begin{figure}
\begin{center}
\includegraphics[scale=.25]{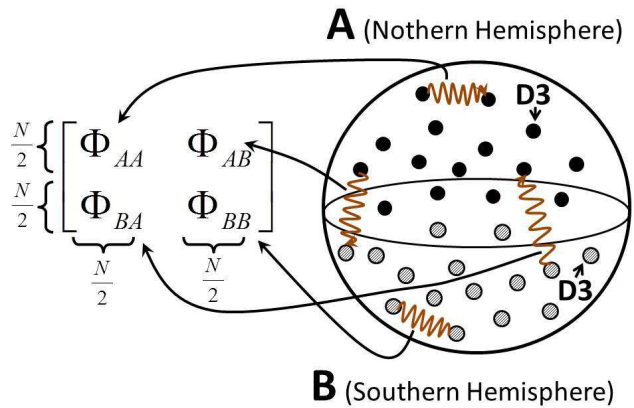}
\caption{A sketch of D3-brane shell background and its dual CFT interpretation. The matrix in the left picture describes the fluctuations around the vacuum. The black (or gray) dots are situated on the surface of S$^5$ sphere in the right picture and denote the $N/2$ D3-branes in the region A (or B), respectively.}
\label{fig:dthree}
\end{center}
\end{figure}

We can obtain a similar volume law result by looking at the D3-brane shell solution \cite{Kraus}.
This corresponds to the setup where D3-branes is distributed at $r=l$ in an $SO(6)$ invariant way. The corresponding supergravity solution is given by
\be
H(r)=\f{R^4}{l^4}\ \ \ (0\leq r\leq l),\ \ \ \ H(r)=\f{R^4}{r^4}\ \ \ (r\geq l). \label{shell}
\ee
Then $S_{ent}$ is computed in the same way by choosing $\gamma$ to be (\ref{ents})
and we find the same result (\ref{enttwod}). In the dual gauge theory, $S_{ent}$ corresponds to the entanglement entropy between the gauge theory defined by $N/2$ D3-branes on the northern hemisphere and the other one defined by $N/2$ D3-branes on the southern hemisphere as in Fig.\ref{fig:dthree}.

\subsection{A Proposal for Generalized Holographic Entanglement Entropy}

The previous holographic consideration leads us to a generalized formulation of holographic entanglement entropy. In general, in a AdS$_{d+1}$/CFT$_{d}$ setup with the classical gravity approximation, the gravity dual is described by a spacetime of the form
M$_{q+d+1}=$Y$^{AdS}_{d+1}\times $ X$_{q}$, where $Y^{AdS}_{d+1}$ denotes a $d+1$ dimensional asymptotically AdS spacetime, while X$_{q}$ does a $q$ dimensional internal compact space.

If we approach the AdS boundary, the total boundary geometry looks like R$^d\times
X_q(\equiv$N$_{q+d}=\de \mbox{M}_{q+d+1})$, where note that the boundary of Y$^{AdS}_{d+1}$ is given by R$^{1,d-1}$ assuming the Poincare coordinate. We define the time coordinate of  R$^{1,d-1}$ to be $t$.
Then we introduce a region $A$ and $B$ so that their boundary $\de A(=\de B)$ divides R$^{1,d-1}\times X_q$ exclusively into $A$ and $B$ at a time $t=t_0$. In other words, $\de A$ is a codimension one surface in R$^{d-1}\times$ X$_q$.

Our main assumption is that this separation of the time slice of the boundary N$_{q+d}$ into two regions $A$ and $B$, corresponds to a factorization of Hilbert space in the dual CFT:
\be
{\mathcal H}_{CFT}={\mathcal H_A}\otimes {\mathcal H_B}. \label{sepw}
\ee
We will discuss the validity of this assumption later. Here we just would like to conjecture that this factorization is always realized for any choice of $A$ and $B$ in CFTs with classical holographic duals, owing to the large $N$ limit.

Under this assumption, we would like to argue that the holographic entanglement entropy for the subsystem $A$ is given by the holographic formula (\ref{hee}),
with $\gamma$ chosen to be the minimal (or extremal in time-dependent cases) surface
whose boundary coincides with that of the region $A$ i.e. $\de \gamma=\de A$.
In particular, when we choose $\de A$ which wraps X$_q$ completely, this prescription is reduced to the standard  holographic entanglement entropy \cite{RT,HRT}. Otherwise, this gravitational entropy gives a new quantity defined in the above. 

General behaviors of such minimal surfaces have been studied in \cite{GrKa}. As shown in that paper, as opposed to minimal surfaces which wrap completely the internal compact manifold X$_q$, minimal surfaces which have non-zero codimension in X$_q$ and extends in a part of AdS directions can approach to only specific surfaces in X$_q$ in the AdS boundary limit $r\to\infty$. For example, when X$_q=$S$^5$, such a surface is only the one given by the equator S$^4$.
This means that we always need to introduce a UV cut off $r_{UV}$ 
in the radial direction of AdS and that
the shape of the minimal surface largely depends on the choice of $r_{UV}$. As we will see later, this difference from usual holographic entanglement entropy in the AdS space is related to the fact that our new entropy generally satisfies the volume law instead of area law.\footnote{We added this paragraph owing to very useful comments from Andreas Karch after this paper appeared on the arXiv.}

If we assume a replica method calculation of entanglement entropy, this prescription is naturally derived from the analysis of generalized gravitational entropy introduced in \cite{LM}. Note also that the pure state property $S_A=S_B$ is obvious and moreover the proof of strong subadditivity and its generalizations can be applied to our generalized minimal surface prescription as in the standard case \cite{SSA,HHM}.

As a non-trivial example, we are interested in the case where $\de A$ is chosen such that the internal space X$_q$ is divided into two regions by $\de A$, while it wraps completely a time slice of Y$^{AdS}_{d+1}$. A particular example of this setup was already discussed in the previous subsection (see the upper left picture in Fig.\ref{fig:geo}).

\subsection{Coincident D3-branes and Entanglement}

As a fundamental example of the generalized holographic entanglement entropy introduced in the previous subsection, we would like to apply this idea to AdS$_5\times$ S$^5$ spacetime. We take the region $A$ (and $B$) to be the northern (and southern) hemisphere of S$^5$ times AdS$_4$. In other words, $\de A$ is defined by (\ref{ents}). In this case, the entanglement entropy reads
\be
S_{ent}=\f{R^2V_3\mbox{Vol}(S^4)}{4G^{(10)}_{N}}\int^{r_{UV}}_0 r^2 dr =\f{4}{9\pi^2}\cdot\f{N^2V_3}{\ep^3}, \label{coind}
\ee
where $r_{UV}\equiv \f{R^2}{\ep}$ is the UV cut off. The infinitesimal parameter
$\ep$ gives the UV cut off (or lattice constant) in the dual CFT, thinking of the expression of the Poincare AdS: $ds^2=R^2\f{dz^2+dx_i^2}{z^2}$ via the coordinate transformation $z=R^2/r$.

As we will argue later, this quantity (\ref{coind}) measures the entanglement entropy between two $SU(N/2)$ ${\mathcal N=4}$ super Yang-Mills, which are interacting with each other within the total system of a $SU(N)$ ${\mathcal N=4}$ super Yang-Mills. Indeed, our holographic result (\ref{coind}) is consistent with the volume law divergence which we found in the field theory calculations for the massless interaction. Geometrically, the origin of volume law divergence is clear because the minimal surface $\gamma$ is
extended to the AdS boundary (refer to the upper left picture in Fig.\ref{fig:geo}).

We can compare this entropy (\ref{coind}) with the maximal possible entropy \cite{SWi} (i.e. the log of the dimension of ${\mathcal H}_A$). The latter is computed as the area of hemisphere of S$^5$ at $r=r_{UV}$:
\be
S_{max}=\f{R^2V_3}{4G^{(10)}_{N}}\cdot\f{r_{UV}^3\mbox{Vol}(S^5)}{2}=\f{1}{2\pi}\cdot \f{N^2V_3}{\ep^3},
\ee
which is indeed larger than (\ref{coind}).

We can generalize this calculation to the cases where the region $A$ is not exactly a half of S$^5$. For example, let us define $A$ such that $\de A$ is given by $\theta=\theta_0$ and $t=0$. Note that as we 
mentioned in the previous subsection owing to the paper \cite{GrKa}, we need to introduce a UV cut off 
$r_{UV}$ to define such a minimal surface and we cannot strictly take $r_{UV}=\infty$ limit. Then $S_{ent}$ should measures the entanglement entropy between $SU(M)$ and
$SU(N-M)$ subsectors of the full $SU(N)$ gauge theory, where $M$ is given by
$M=\f{8N}{3\pi}\cdot \int^{\theta_0}_0 d\theta\sin^4\theta$, being proportional to the volume of region $A$.

\begin{figure}
\includegraphics[scale=.5]{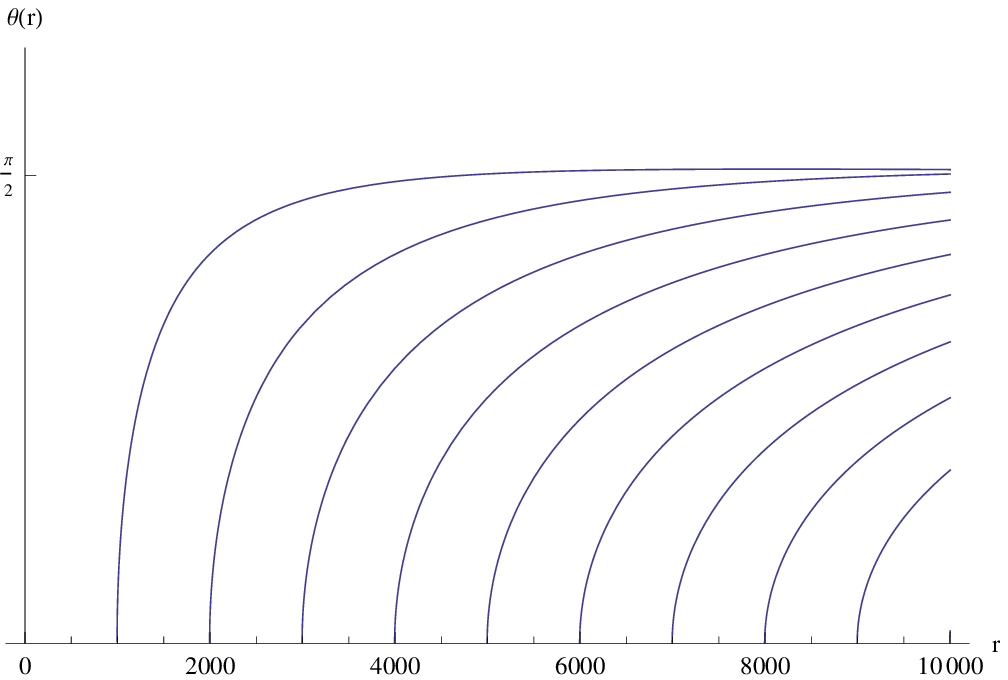}
\includegraphics[scale=.5]{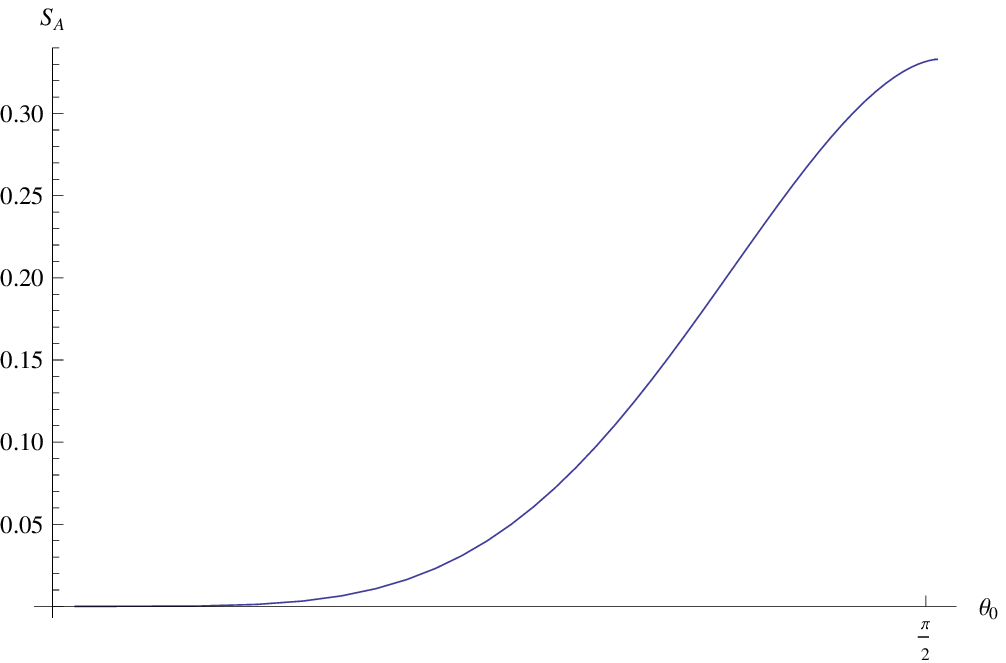}
\includegraphics[scale=.5]{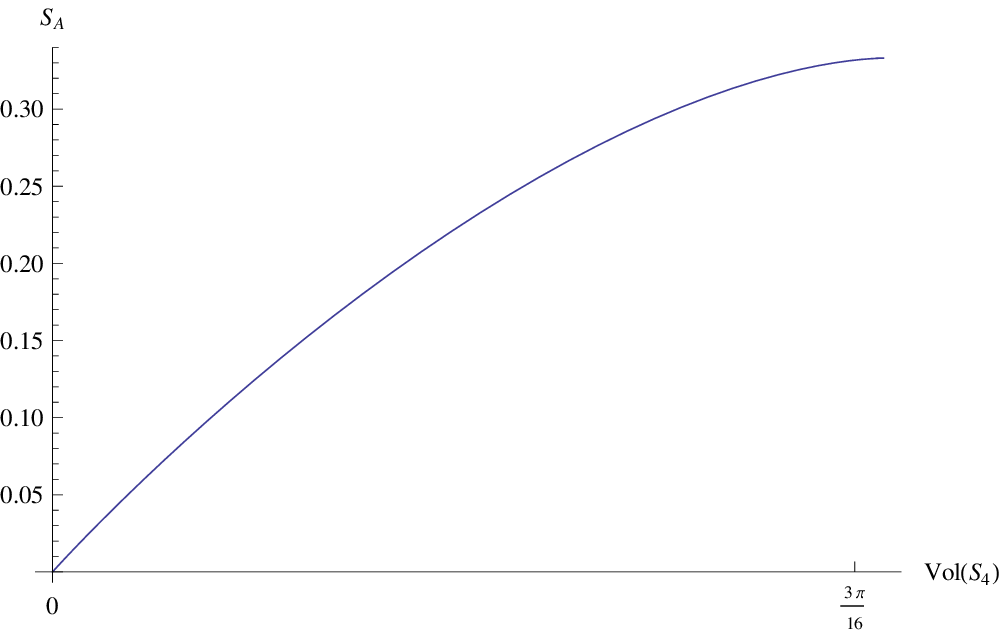}
\caption{Left: Profile of minimal surfaces $\theta(r)$ in the bulk. Middle: holographic entanglement entropy $S_{ent}$ as a function of $\theta_0$. Right: $S_{ent}$ versus the (normalized) volume of the region $A$ in S$^5$, given by $\int^{\theta_0}_0d\theta\sin^4\theta$. Note that we took $r_{UV}=10000$ in all graphs. In the middle and right graph, we plotted only the integral of (\ref{aligg}) in the unit of $r_{UV}^3$ as $S_{ent}$.}
\label{fig:theta}
\end{figure}

In this case, the minimal surface is described by a profile
\be
t=0,\ \ \ \theta=\theta(r).
\ee
The function $\theta(r)$ can be found by minimizing the holographic entanglement entropy functional:
\ba
S_{ent}=\f{R^2V_3\mathrm{Vol}(S^4)}{4G^{(10)}_N}
\int_{r_0}^{r_{UV}}r^2\sin^4\theta(r)\sqrt{1+r^2\dot{\theta}^2(r)}dr, \label{aligg}
\ea
which leads to the following equation
\ba\label{ms}
\p_r\left[r^2\sin^4\theta(r)\f{r^2\dot{\theta}(r)}{\sqrt{1+r^2\dot{\theta}^2(r)}}\right]
=4\cos\theta(r)\sin^3\theta(r)r^2\sqrt{1+r^2\dot{\theta}^2(r)},
\ea
and by imposing the boundary condition $\theta(r_{UV})=\theta_0$. The constant $r_0$ is defined by the value of $r$ at the turning point i.e. $|\theta(r_0)|=\infty$.
The profile of minimal surfaces have been found numerically and we plotted this in the left picture of Fig.\ref{fig:theta}. The holographic entanglement entropy is plotted in the middle picture of the same figure. Especially in the right graph, we plotted the entropy $S_{ent}$ as a function of the volume of the region $A$. We can find that for small $\theta_0$, the entropy is proportional to the volume of $A$. Therefore it satisfies the volume law instead of the area law as typical in highly non-local field theories (see \cite{ShTa} for such examples of volume law).

Next, to study a finite temperature example, let us replace the AdS$_5$ with the AdS$_5$ Schwarzschild black hole:
\ba
&& ds^2=R^2\left[-\f{f(z)}{z^2}dt^2+\f{dz^2}{f(z)z^2}+\f{dx_i^2}{z^2}\right]+
R^2\left(d\theta^2+\sin^2\theta d\Omega_4^2\right), \no
&& f(z)=1-\left(\f{z}{z_H}\right)^4,
\ea
where the black hole temperature $T$ is given by $T=\f{1}{\pi z_H}$.

The holographic entanglement entropy is computed as follows
\ba
S_{ent}&=&\f{2\pi^2 R^8V_3}{3G_N}\cdot \int^{z_H}_\ep \f{dz}{z^4\s{f(z)}},\no
&=& \f{4N^2V_3}{9\pi^2\ep^3}+\f{4\pi \ap_0 V_3N^2}{9}T^3.
\ea
where $\ap_0=\f{\s{\pi}\Gamma(5/4)}{\Gamma(3/4)}\simeq  1.31$.

Note that the finite term in $S_{ent}$
\be
S^{finite}_{ent}=\f{4\pi\ap_0 V_3N^2}{9}T^3\simeq 1.83\cdot  V_3 N^2T^3,
\ee
looks similar to the thermal entropy up to a numerical constant, as expected. This finite part is expected to approximately describe the entanglement between two $SU(N/2)$ ${\mathcal N}=4$ super Yang-Mills theories due to thermal effects. Indeed, we can compare this with the thermal entropy of $SU(N/2)$ ${\mathcal N}=4$ super Yang-Mills at zero coupling
\be
S^{free}_{thermal}=\f{\pi^2}{6}V_3 N^2T^3\simeq 1.64\cdot V_3 N^2T^3,
\ee
with a good semi-quantitative agreement as analogous to the black 3-brane entropy \cite{GKPB}.

As the final example, we would like to come back to the D3-brane shell solution (\ref{shell}),
corresponding to a state in the Coulomb branch (see Fig.\ref{fig:dthree}). In the current setup we take the UV cut off $r_{UV}=R^2/\ep$ to be much larger than the mass scale $M$. The holographic entanglement entropy is computed as follows:
\ba
S_{ent}&=&\f{V_3\cdot \mbox{Vol}(S^4)}{4G_N}\int^{r_{\infty}}_{0}dr r^4 \s{H(r)} \no
&=&\f{4V_3N^2}{3\pi^2}\left[\f{1}{3\ep^3}-\f{2}{15}\cdot\f{l^3}{R^6}\right] \no
&=&\f{4V_3N^2}{3\pi^2}\left[\f{1}{3\ep^3}-\f{1}{60}\cdot\f{\pi^{3/2}M^3}{\lambda^{3/2}}\right].
\ea
This entropy is a monotonically decreasing function of $M$ as expected, because the presence of mass clearly reduces quantum entanglement.

\subsection{Comments on Factorization of Hilbert Space}

Finally we would like to come back to the issue on how we can realize the factorization of Hilbert space (\ref{sepw}). When the surface $\de A$ divides the real space R$^{d-1}$ where the CFT is defined on, this factorization can be explicitly realized by e.g. discretizing the CFT on a lattice. On the other hand, when $\de A$ divides the internal space X$_q$ into two subregions, with $\de A$ wrapped on $R^{d-1}$ completely, the factorization is not obvious.
This is closely related to a deep question in AdS/CFT how the geometry of internal space appears from the data of CFT (see e.g. \cite{Beren}). We are not going to give a complete answer to this profound question in this paper. Instead, we will give some heuristic arguments which support such a  factorization, which motivates us to conjecture the factorization in the generalized cases.

First, if we remember the Coulomb branch setup (\ref{col}), it is clear that we will approximately have the factorization (\ref{sepw}) if we take the cut off scale $\Lambda$ much smaller than the mass $M$ of bi-fundamental fields. In this situation, ${\mathcal H_A}$ and ${\mathcal H_B}$ are the two Hilbert spaces constructed from excitations in each of the two $SU(N/2)$ super Yang-Mills theory. It is curious that in this case we are looking at essentially a flat space rather than the AdS space as is clear from the shell metric (\ref{shell}).

However, once we take the energy scale to be larger than the mass scale, then we cannot
regard the system as two $SU(N/2)$ gauge theories but we should treat them as an interacting $SU(N)$ gauge theory. Therefore we need to take into account bi-fundamental fields directly as active fields in the factorization (refer to Fig.\ref{fig:dthree}).

Thinking of the D3-brane shell solution, we can identify the positions of each of $N$ D3-branes. They are described by the eigenvalues $\vec{\vp}_a=(\vp^1_{a},\vp^2_a,\ddd , \vp^6_a)$ of the
six $N\times N$ matrices $\vec{\Phi}_{ab}=(\Phi^1_{ab},\Phi^2_{ab},\ddd,\Phi^6_{ab})$, where
$a,b=1,2,\ddd,N$. For the non-diagonal components of the matrices $\vec{\Phi}_{ab}$ with $a\neq b$ we can, for example, naturally assign the middle position $\f{1}{2}(\vec{\vp}_a+\vec{\vp}_b)$.
In this way, there is a way to assign all of $N\times N$ components specified by the pair $(a,b)$ to $N^2$ different positions on R$^6$. Therefore we can project them in the radial direction and have a direct map between $N^2$ elements $(a,b)$ and $N^2$ position on S$^5$. In other words, a matrix field $\Phi_{ab}$ with both $a$ and $b$ belong to the first
(or second) $SU(N/2)$ indices is regarded as a vector in ${\mathcal H_A}$ (or ${\mathcal H_B}$). When $a$ belongs to the first one and $b$ does to the second one, then we can decide whether it is regarded as a vector in  ${\mathcal H_A}$ or ${\mathcal H_B}$ by examining whether the vector $\f{1}{2}(\vec{\vp}_a+\vec{\vp}_b)$ belong to the south or north hemisphere. Clear this separation is just an example and we can think of many other definition of the factorization of Hilbert space (\ref{sepw}). This problem has some similarities with the ambiguity which occurs in a precise definition of more standard entanglement entropy in gauge theories \cite{gauge,lattice} even when we define the subsystem $A$ and $B$ by a geometrical separation of the real space.

 Owing to the large $N$ limit, once we admit the previous rule, the $N^2$ point is expected to densely cover S$^5$. Thus we may approximately regard the matrices $\delta\vec{\Phi}_{ab}$, which describe the fluctuations around the classical expectation value  $\vec{\vp}_a$,
as six functions on S$^5$:
\be
\delta\vec{\Phi}_{ab}\to \delta\vec{\Phi}(\Omega).   \label{map}
\ee

In this way, we can regard the originally four dimensional gauge theory as an effectively nine dimensional field theory on R$^{1,3}\times $S$^5$. However, note that the S$^5$ direction is not standard in that it does not have a conventional kinetic term and should be highly non-local. This fact can be seen from the volume law which we observed in the holographic calculation plotted in Fig.\ref{fig:theta}.
Nevertheless, (\ref{map}) allows us to divide the total Hilbert space into the factorized form
(\ref{sepw}) as we wanted to show. See \cite{ShTa} for a calculation of entanglement entropy in non-local field theories.

It is also straightforward to extend these arguments to more general Coulomb branch solutions.
In particular, if we take $M\to 0$ limit of the D3-brane shell, then we can obtain the AdS$_5\times$ S$^5$ case. It is also possible to extend our argument to the global AdS$_5$. As argued in \cite{Beren}, we will again get a shell like configuration in the large $N$ limit, where the eigenvalues of the transverse scalars are distributed on a S$^5$ in R$^6$. Therefore we can apply the same argument as the previous one for the shell solution.

\section{Conclusions and Discussions}\label{fink}

In this paper we studied the quantum entanglement between two CFTs which are interacting each other. Especially we computed (Renyi) entanglement entropy $S_{ent}$ from the viewpoint of both field theory and AdS/CFT.

First, we analytically computed the (Renyi) entanglement entropy between two free scalar field theories in $d$ spacetime dimensions, which are interacting with each other in several ways. We considered massless (marginal) and massive (relevant) interactions and moreover analyzed time-dependent examples.
We provided two different but equivalent formulations for this problem: the real time formalism using the wave functional and the Euclidean replica formalism using boundary states. We confirmed that they give the same results in several examples.

These computations show that the entanglement entropy between the two CFTs in the presence of massless interactions follows the volume law $S_{ent}\propto V_{d-1}\Lambda^{d-1}$ in a universal way, where $\Lambda$ is the UV cut off. When they are interacting via massive interactions, the entanglement entropy is suppressed and we fund that the entropy becomes finite for $d\leq 4$ and for $d\geq 5$ it behaves as $S_{ent}\propto V_{d-1}\Lambda^{d-5}\log \Lambda$. This is expected because the massive interaction is suppressed in the high energy region.

We also studied the time evolution of entanglement entropy $S_{ent}$ between a copy of two dimensional CFTs when we turn on the mutual interactions suddenly at a time. Our field theory results show that in the case of massless interactions, $S_{ent}$ increases almost instantaneously and saturates to a constant value, which has again the volume law divergence. On the other hand, in the case of massive interactions, $S_{ent}$ increases slowly as $S_{ent}\propto t^2$, where $S_{ent}$ takes always finite values.

In the final part of this paper, we proposed a holographic dual calculation of entanglement entropy between the two interacting CFTs. This is done by generalizing the holographic entanglement entropy so that we divide the internal space, such as the S$^5$ in the AdS$_5\times$ S$^5$ type IIB solution, into two subregions $A$ and $B$. We explicitly studied the four dimensional $SU(N)$ ${\mathcal N}=4$ super Yang-Mills and computed the entanglement entropy between two $SU(N/2)$ sub-sectors, employing the Coulomb branch solutions. This measures the entanglement between $N/2$ D3-branes on the northern hemisphere of S$^5$ and the other $N/2$ D3-branes on the southern hermisphere. Motivated by these we conjectured a generalized holographic entanglement entropy, where we can choose the entangling surface $\gamma$ as any minimal (or extremal) surfaces in an AdS$_{d+1}\times$ X$_q$ spacetime of ten or eleven dimensional supergravities. However, we left an important of problem of consistency between the proposed factorizations of Hilbert space and the gauge invariance of total system for future problem, though we gave several supporting arguments.

Our generalization of holographic entanglement entropy is closely to related to a basic question of AdS/CFT: how the S$^5$ geometry emerges from the ${\mathcal N}=4$ super Yang-Mills theory. The detailed understanding of precise factorizations of Hilbert spaces, which we left for a future problem, may be an important clue for this problem. One useful approach may be to use the idea of the multi-scale entanglement renormalization ansatz (MERA) \cite{Vidal,Hae} as this offers a manifest geometrization of wave functions of quantum many -body systems or field theories, conjectured to be equivalent to the AdS/CFT \cite{MERA,BHMERA,cMERA}.

Notice that we provided such holographic computations only when the two CFTs are interacting via massless interactions. In the case of massive interactions, field theory results argue that the UV divergence is suppressed a lot. This suggests that
its gravity dual looks like the upper right picture in Fig.\ref{fig:geo}, where the minimal surface $\gamma$ does not reach the AdS boundary. However, we should be careful that if we naively consider such a geometry with two boundaries which are causally connected, it contradicts with the topological censorship \cite{top}. Thus we need to take into account quantum corrections or complicated structures of internal manifolds. One concrete example may be the one \cite{Massive} which employs the massive gravitons. We would like to leave a detailed holographic study of two entangling CFTs with massive interactions as an interesting future problem. It will also be intriguing to consider a holographic construction of entangling time-dependent CFTs as we analyzed in the field theory calculations.

\section*{Acknowledgements}

We would like to thank Song He, Masahiro Nozaki, Masaki Shigemori and Tomonori Ugajin for useful discussions. We also would like to thank
Andreas Karch for giving us very useful comments after this paper appeared on the net. AM is also grateful to Mohsen Alishahiha for his supports.
NS and TT are supported by JSPS Grant-in-Aid for Scientific Research (B) No.25287058. TT is also supported by JSPS Grant-in-Aid for Challenging
Exploratory Research No.24654057. AM is supported by Iran Ministry of Science, Research, and Technology grant for PhD students sabbatical.
TT is also supported by World Premier International Research Center Initiative (WPI Initiative) from the Japan Ministry
of Education, Culture, Sports, Science and Technology (MEXT).


\end{document}